\documentclass[lettersize,journal]{IEEEtran}
\usepackage{amsmath}
\usepackage{amsfonts}
\usepackage{mathtools}
\usepackage{algorithmic}
\usepackage{algorithm}
\usepackage{array}
\usepackage{pifont}
\usepackage{xcolor}
\usepackage{tabularx}
\usepackage{makecell}
\usepackage{multirow}
\usepackage[compact]{titlesec}
\usepackage[caption=false,font=normalsize,labelfont=sf,textfont=sf]{subfig}
\usepackage{textcomp}
\usepackage{stfloats}
\usepackage{url}
\usepackage{verbatim}
\usepackage{graphicx}
\usepackage{cite}
\usepackage[hidelinks]{hyperref}
\hypersetup{
  colorlinks = true, 
  citecolor = black, 
  linkcolor = black, 
  urlcolor = black
}

\newcommand{\RNum}[1]{\uppercase\expandafter{\romannumeral #1\relax}}
\DeclareMathOperator*{\argmax}{argmax}

\newcommand{\rw}[1]{\textcolor{black}{#1}}

\begin{document}

\title{Joint Semantic Transfer Network for IoT Intrusion Detection}

\author{Jiashu Wu,
        Yang Wang,~\IEEEmembership{Member,~IEEE}\thanks{* Yang Wang is the corresponding author},
        Binhui Xie, 
        Shuang Li, 
        Hao Dai, 
        Kejiang Ye,~\IEEEmembership{Member,~IEEE},
        and Chengzhong Xu,~\IEEEmembership{Fellow,~IEEE}% <-this % stops a space
\thanks{Jiashu Wu, Hao Dai, Yang Wang and Kejiang Ye are with Shenzhen Institute of Advanced Technology, Chinese Academy of Sciences, Shenzhen 518055, China. Email: \{js.wu, hao.dai, yang.wang1, kj.ye\}@siat.ac.cn}% <-this % stops a space
\thanks{Jiashu Wu and Hao Dai are also with University of Chinese Academy of Sciences, Beijing 100049, China. Email: \{wujiashu21, daihao19\}@mails.ucas.ac.cn}% <-this % stops a space
\thanks{Binhui Xie and Shuang Li are with Beijing Institute of Technology, Beijing 100081, China. Email: \{binhuixie, shuangli\}@bit.edu.cn}% <-this % stops a space
\thanks{Chengzhong Xu is with the State Key Laboratory of IoT for Smart City, Faculty of Science and Technology, University of Macau, Macau 999078, China. Email: czxu@um.edu.mo}% <-this % stops a space
\thanks{Portion of this paper \cite{10.1145/3394171.3413995} has been published in the 28th ACM International Conference on Multimedia (ACM MM'2020), 12-16 October, 2020. DOI: https://doi.org/10.1145/3394171.3413995}% <-this % stops a space
\thanks{Manuscript received January 00, 2022; revised January 00, 2022.}
\thanks{Copyright (c) 2022 IEEE. Personal use of this material is permitted. However, permission to use this material for any other purposes must be obtained from the IEEE by sending a request to pubs-permissions@ieee.org. }}

\markboth{IEEE Internet of Things Journal,~Vol.~00, No.~0, August~2022}%
{Shell \MakeLowercase{\textit{et al.}}: A Sample Article Using IEEEtran.cls for IEEE Journals}


\maketitle

\begin{abstract}
In this paper, we propose a Joint Semantic Transfer Network (JSTN) towards effective intrusion detection for large-scale scarcely labelled IoT domain. As a multi-source heterogeneous domain adaptation \rw{(MS-HDA) method}, the JSTN integrates a knowledge rich network intrusion (NI) domain and another small-scale IoT intrusion (II) domain as source domains, and preserves intrinsic semantic properties to assist target II domain intrusion detection. The JSTN jointly transfers the following three semantics to learn a domain-invariant and discriminative feature representation. The scenario semantic endows source NI and II domain \rw{with characteristics from} each other to ease the knowledge transfer process via a confused domain discriminator and categorical distribution knowledge preservation. It also reduces the source-target discrepancy to make the shared feature space domain-invariant. Meanwhile, the weighted implicit semantic transfer boosts discriminability via a fine-grained knowledge preservation, which \rw{transfers the} source categorical distribution to the target domain. The source-target divergence guides the importance weighting during knowledge preservation to reflect the degree of knowledge learning. Additionally, the hierarchical explicit semantic alignment performs centroid-level and representative-level alignment with the help of a geometric similarity-aware pseudo-label refiner, which exploits the value of unlabelled target II domain and explicitly aligns feature representations from a global and local perspective in a concentrated manner. Comprehensive experiments on various tasks verify the superiority of the JSTN against state-of-the-art comparing methods, on average a 10.3\% of accuracy boost is achieved. \rw{The statistical soundness of each constituting component and the computational efficiency are also verified. }
\end{abstract}


\section{Introduction}\label{sec:sec1_introduction}

\IEEEPARstart{A}{s} the Internet of Things (IoT) devices become more ubiquitous in our daily life \cite{chen2014vision_iot_prevalent,iot_application_agriculture,iot_application_smartgrid}, they have transformed various fields such as healthcare \cite{iot_application_healthcare1,iot_application_healthcare2} and public transport \cite{iot_application_public_transport} into a smart space. However, IoT infrastructures are usually formed by resource-limited devices with infrequent security maintenance effort from their vendors \cite{eskandari2020passban}, which poses security threats for malicious attacks to take advantage of IoT security flaws and perform intrusions which harm the underlying IoT infrastructures \cite{iot_security1,iot_security2} and the applications they support. Therefore, a robust intrusion detection system (IDS) \cite{restuccia2018securing} is crucial to effectively detect these malicious intrusions faced by IoT infrastructures. 

With the rapid development of machine learning (ML) and deep learning (DL) techniques, recently, several DL-based \rw{IDSs become popular}. For instance, Anthi et al., \cite{anthi2019smart_home_iot} analysed the IDS performance of several supervised methods under the smart home IoT scenario, such as Naive Bayes classifier, Support Vector Machine, etc. The results verified the intrusion detection (ID) effectiveness of these methods. However, these methods highly depend on a vast amount of fully labelled data, which is expensive to collect and labourious to annotate. This is particularly difficult for IoT intrusion (II) detection, since data generated by IoT devices usually involves user privacy issues \cite{benkhelifa2018critical,abdelmoumin2022survey}, \rw{which hinder the} publication of IoT intrusion detection data. Besides, these ML models are less capable of handling newly emerged intrusion types due to the shortage of annotated data. Considering that intrusion detection data for IoT is expensive to collect and seldom available, several domain adaptation (DA) approaches were proposed to transfer the rich knowledge from network intrusion (NI) domain to facilitate the intrusion detection for label-scarce IoT domains. Since the NI data is relatively richer than II domains \cite{booij2021ton_iot}, these DA approaches treated the NI as the source domain, and the II as the target domain. As the network and IoT share several common attack types, by mapping both domains into a common feature subspace, these DA approaches can transfer the enriched NI knowledge to assist intrusion detection in the target IoT domain. For instance, Vu et al., \cite{vu2020deep_autoencoder} utilised two autoencoders as feature extractors for source and target domain, and minimised the maximum mean discrepancy (MMD) between their bottleneck layers to achieve knowledge transfer. However, previous DA-based ID models usually produced coarse-grained alignment. They aligned the source and target domain into a common feature subspace by brute force without transferring intrinsic semantic properties, which may result in instances from different categories being confounded together and therefore hurt the discriminability of learned features. 

To address the limitations of coarse-grained DA-based ID models and facilitate better transferability, in this paper, we propose a Joint Semantic Transfer Network (JSTN) which leverages the intrinsic semantic knowledge between domains to facilitate a more fine-grained knowledge transfer. Considering that there is a huge domain gap between the source NI and target II domain due to heterogeneities such as different feature representations, different distributions, etc., the effectiveness of direct domain adaptation may be hindered. Hence, to ease the adaptation process, we utilise another labelled II domain as source domain, which is smaller in scale than both the source NI and target II domain, to partially mask the heterogeneities. All domains have their own domain encoders to map instances into a common feature subspace, and a domain discriminator is confused to shorten the source NI - source II divergence and the source-target divergence. The predicted categorical distribution knowledge is also transferred between source domains for better discriminability. By introducing this auxiliary small-scale source II domain, it can equip the heterogeneous source NI domain with the semantics of IoT scenarios. By drawing the network and IoT \rw{intrusion scenarios} closer and letting them become similar to partially mask scenario heterogeneities, the source NI and II domain form a holistic source domain with rich intrusion knowledge and IoT scenario characteristics, which can therefore benefit the source-target knowledge transfer performed later. 

Additionally, to overcome the category confounding caused by the coarse-grained feature alignment, we propose a weighted implicit semantic transfer, which preserves the correlation knowledge between categories from the source to the target domain. It is intuitive that the same class from either source or target domain should share a relatively similar categorical distribution. During weighted implicit semantic transfer, the knowledge from the source NI and II domain are weighted based on their divergence with the target domain to dynamically emphasise varied source domain importance which reflects the degree of knowledge learning. The weighted implicit semantic alignment can effectively enhance the discriminability of the learned feature. 

Given that the majority of target II domain instances are unlabelled, while exploring unlabelled target data is beneficial during domain adaptation \cite{li2013learning,10.1145/3394171.3413995}, especially when there are huge heterogeneities present between domains. Therefore, we propose a hierarchical explicit semantic alignment from centroid-level and representative-level. The centroid-level alignment matches each category between the source domain, the target domain, and the combination of source and target domain from a global centroid perspective. Considering that only utilising the global centroid-level alignment may hurt the concentration of aligned features, we also leverage the representative-level alignment, which performs a class-wise representative selection and minimises the pairwise divergence between class-wise representatives from source and target domain. Hence, it boosts the concentration of aligned features yielded by the semantic alignment from a local perspective without causing heavy computational burden. To fully excavate the potentials of unlabelled target II data during hierarchical explicit semantic transfer and avoid the misleading direct pseudo-label assignment \cite{tsai2016learning,hsieh2016recognizing}, a pseudo-label refiner (PLR) is leveraged to assign unlabelled target II instances with pseudo-labels via an ensemble approach. It will investigate the geometric similarity between each unlabelled target instance and the centroid of labelled instances of each category, and regard the most geometrically similar category as the geometric label. Then, the pseudo-label refiner will only assign pseudo-label to an instance if the geometric label agrees with the prediction yielded by the shared classifier. Assisted by the more accurate pseudo-label refiner, the hierarchical explicit semantic alignment with a global and local perspective can explicitly minimise domain divergence in a concentrated manner and promote discriminability. 

Ultimately, by jointly utilising these semantics, the JSTN model can robustly transfer enriched knowledge from the knowledge rich NI domain and a small-scale II domain to facilitate more accurate intrusion detection of the scarcely-labelled target II domain and hence secure the IoT infrastructures. 

In summary, the contributions of this paper are as follows:
\begin{itemize}
  \item We utilise the joint semantic transfer to leverage the enriched knowledge of NI domain with the help of an auxiliary small-scale II domain to facilitate more accurate intrusion detection of the large-scale scarcely-labelled target II domain. 
  \item We propose a novel Joint Semantic Transfer Network (JSTN) that explores and excavates the semantic transfer to achieve a more effective intrusion knowledge transfer despite significant heterogeneities present between NI and II domains. 
  \item We conduct comprehensive experiments of several tasks on $5$ well-known intrusion detection datasets and demonstrate the effectiveness of the JSTN algorithm, exceeding state-of-the-art comparing methods. 
\end{itemize}

The rest of the paper is organised as follows: Section \ref{sec:section_related_work} summarises related works on signature-based, ML-based and DA-based ID approaches, explains their limitations and reveals our research opportunities. \rw{Section \ref{sec:section_model_and_the_jstn_architecture} presents the model and the architecture of the JSTN method. The details of the proposed JSTN method are presented in Section \ref{sec:section_jstn_algorithm_details}. }The experimental setup, results and insight analyses are given in Section \ref{sec:section_experiment}. Section \ref{sec:section_conclusion} concludes the paper. 

\section{Related Work}\label{sec:section_related_work}

\subsection{Signature-based Intrusion Detection}\label{sec:section_signature_based_intrusion_detection}

As a popular research direction, several signature-based intrusion detection methods have been proposed. They maintained a set of signatures or rules of malicious attacks and performed intrusion detection by matching incoming network traffic with these pre-defined attack patterns. Zhang et al., \cite{zhang2015communication} proposed a preventive measure specifically targeting DDoS attacks on IoT devices. It kept track of the content of incoming requests. If requests from a node show a pattern, e.g., similar meaningless content being repeatedly sent, the preventive measure will flag the corresponding sender as malicious and refuse its future requests subsequently. Dietz et al., \cite{dietz2018iot} proactively performed an automatic scan of neighbouring IoT devices for potential vulnerabilities such as using default credential settings. Once a vulnerable IoT device is detected, it will be temporarily isolated since these IoT devices suffer from a higher chance to be compromised and be manipulated as part of the malicious Botnet. Chen et al., \cite{jun2014design} utilised complex event processing (CEP) technique, a technique to filter and process real-time events. The CEP required a pre-defined rule pattern repository which contains rules and patterns of common IoT security violations. Summerville et al., \cite{summerville2015ultra} presented a lightweight deep packet anomaly detection strategy via efficient bit-pattern matching. The patterns of the payload contents were studied, and the n-gram matching algorithm was leveraged to find pattern matching in an efficient manner. 

Although these previous signature-based ID methods can produce satisfying results, they require substantial expert knowledge to build the pattern repository as the working foundation. The expert knowledge is usually labourious to acquire, barely thorough and complete, and is unable to tackle newly emerged attack types if the pattern repository is not updated on a frequent basis. Hence, it leaves rooms for other research directions. 

\subsection{Machine learning-based Intrusion Detection}\label{sec:section_machine_learning_based_intrusion_detection}

Machine learning and deep learning techniques can also be applied to tackle intrusion detection for IoT scenarios. Shukla \cite{shukla2017ml} presented several new intrusion detection methods based on K-means clustering, decision tree, and an ensemble of these two classical ML algorithms. The proposed approach is lightweight, and is capable of accurately detecting wormhole attacks targeting IoT under 6LoWPAN network environment. Anthi et al., \cite{anthi2019smart_home_iot} focused on the intrusion detection of smart home IoT devices. Several popular classifiers such as Naive Bayes, support vector machines, etc., were evaluated to detect $4$ mean network attack categories on a realistic testbed. Ge et al., \cite{ge2019deep}, McDermott et al., \cite{mcdermott2018botnet} and Meidan et al., \cite{meidan2018} all focused on leveraging deep learning-based methods. A feedforward neural network, bidirectional-LSTM recurrent neural network and a deep autoencoder were constructed to perform intrusion detection for IoT devices, respectively, and demonstrated satisfying outcomes. 

However, these ML and DL-based methods require a large-scale labelled dataset, which is expensive and labourious to acquire. Some datasets become out-of-date quickly as IoT \rw{devices and }attacks keep evolving, which hinder the effectiveness of these methods. Therefore, it naturally leads to the domain adaptation-based (DA) methods \cite{li2018domain,li2021generalized}, which performs knowledge transfer to facilitate intrusion detection of data-scarce IoT spaces. 

\subsection{Domain Adaptation and its application in Intrusion Detection}\label{sec:section_domain_adaptation_and_its_application_in_intrusion_detection}

\textbf{Heterogeneous Domain Adaptation} HDA transfers knowledge from a knowledge rich domain to facilitate learning in a similar but knowledge scarce target domain. The source and target domain \rw{present heterogeneities. }For instance, intrusion data from the network and IoT domain can have different types of devices that work under different environments, using different feature sets, and follow different distributions, etc. Several research efforts have been presented to address the HDA problem with specific focus on the feature-level. Wang et al., \cite{wang2011heterogeneous} utilised manifold alignment (DAMA) to construct mappings to project source and target data to a latent space while preserving the label topology. Hoffman et al., \cite{hoffman2013efficient} presented max-margin domain transforms (MMDT) to simultaneously learn the feature projection and the classifier. Chen et al., \cite{chen2016transfer} proposed the transfer neural tree (TNT) algorithm with stochastic pruning to perform feature transformation and enhance prediction accuracy. Yao et al., \cite{yao2020discriminative} proposed the discriminative distribution alignment (DDA) that incorporated several losses such as cross-entropy loss (DDAC) and squared loss (DDAS) to improve the data separability during alignment. However, these methods \rw{mainly focused} on the feature-level information, none of them \rw{leveraged the} intrinsic semantic correlations contained in the predicted distributions, which may result in confounded attack types and confused predictions if work on IoT intrusion detection. 

Besides, some research efforts tackled the HDA problem by explicitly enforcing domain alignment. Tsai et al., \cite{tsai2016learning} presented cross-domain landmark selection (CDLS) to learn a domain-invariant feature subspace for HDA via cross-domain landmarks. To jointly match the marginal and class-conditional distributions, Hsieh et al., \cite{hsieh2016recognizing} presented the generalized joint distribution adaptation (G-JDA) method and confirmed its effectiveness. To circumvent the negative effect brought by falsely-assigned pseudo-labels, Yao et al., \cite{yao2019heterogeneous} proposed the soft transfer network (STN) which utilised soft labels during alignment. However, some of these methods \rw{directly used the} predicted label as pseudo-label, which will cause severe negative transfer due to falsely-assigned pseudo-labels. These wrong labels will mislead the model, the model will then accumulate more wrong labels, which forms a negative loop. Although methods such as STN attempted to avoid the negative transfer incurred by wrongly-assigned pseudo-labels, they failed to consider the intrinsic geometric semantic contained in the feature space, which can effectively guide the pseudo-label assignment and boost the pseudo-label confidence. 

Finally, considering that utilising two source domains may enhance the knowledge transfer process, Yao et al., \cite{yao2021multisource} proposed a conditional weighting adversarial network (CWAN) to address the multi-source HDA problem. However, it did not verify the effectiveness of multi-source DA method on intrusion detection tasks, which left a void to be filled. Hence it did not attempt the idea of scenario semantic to be used between network and IoT domains either, and also lacked the joint consideration of semantic transfer. 

\textbf{Domain Adaptation-based Intrusion Detection} The capability of the DA to transfer knowledge and facilitate robust learning in the target domain makes it a perfect choice for intrusion detection. Vu et al., \cite{vu2020deep_autoencoder} trained two autoencoders for a label rich and a label scarce IoT domain separately, and \rw{bridged the gaps} between the bottleneck layers of these two autoencoders by minimising the maximum mean discrepancy (MMD). Hu et al., \cite{hu2022deep} proposed a deep subdomain adaptation network with attention mechanism (DSAN-AT), which utilised the local MMD to boost the prediction accuracy \rw{and an} attention mechanism \rw{to prevent} overly long convergence time. To circumvent the labour-intensive dataset collection process, Ning et al., \cite{ning2021malware} proposed a knowledge transfer (KT) ConvLaddernet to work under a semi-supervised setting, i.e., transfer knowledge from a small-scale source domain to facilitate intrusion detection of the target domain. Although previous DA-based methods have been applied to perform intrusion detection, they failed to jointly consider the implicit categorical and explicit distance semantics during knowledge transfer, which may hinder their effectiveness. Besides, these DA-based methods did not realise that utilising a network intrusion domain plus a small-scale IoT intrusion domain can boost the intrusion performance of a large-scale scarcely-labelled target IoT domain. Hence, they \rw{left the} potential of scenario semantic untouched. 

\begin{figure*}[t]
  \begin{center}
    \includegraphics[width=0.97\textwidth,keepaspectratio]{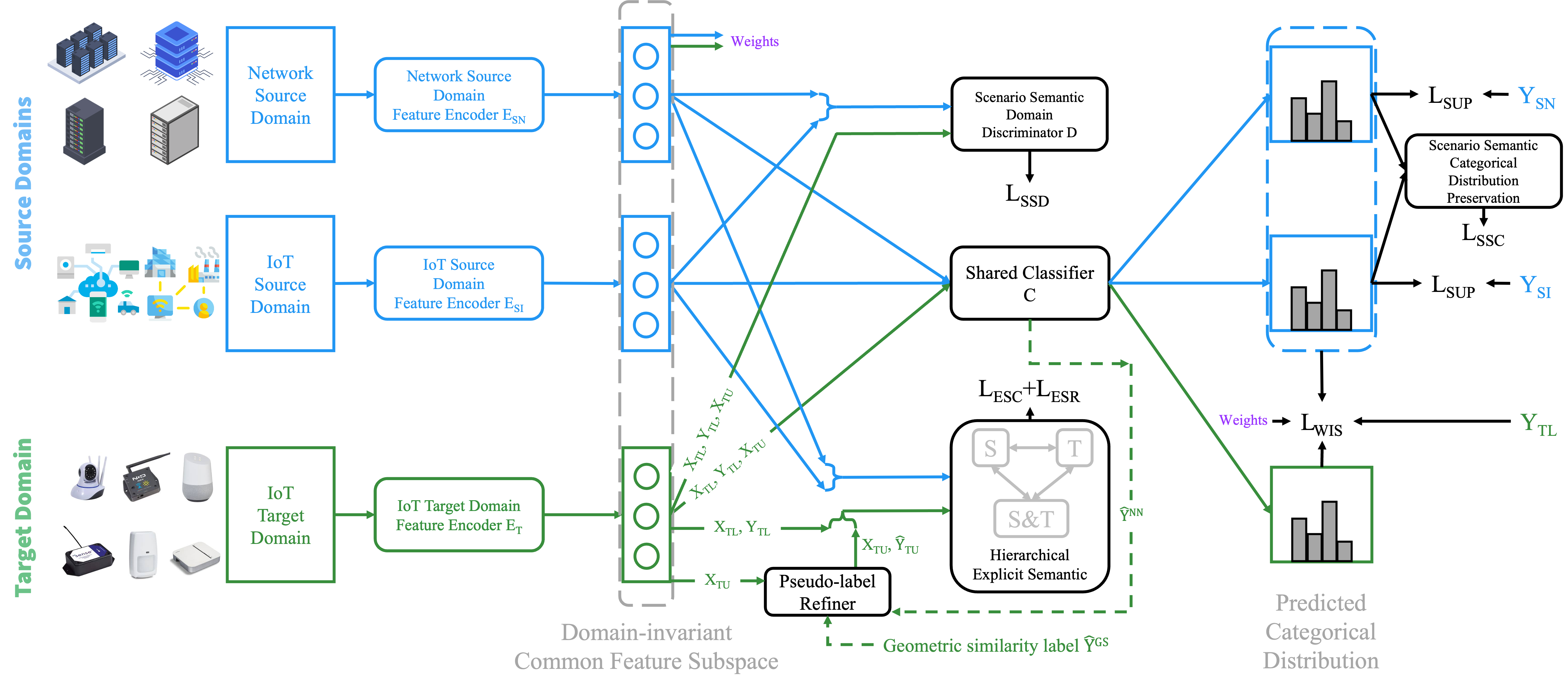}\\
    \caption{The architecture of the JSTN model. }
    \vspace{-0.5cm}
    \label{fig:figure_network_framework}
  \end{center}
\end{figure*}

\section{\rw{Model and JSTN Architecture}}\label{sec:section_model_and_the_jstn_architecture}

\rw{In this section, we will mainly present the problem setting, followed by the architecture of the JSTN algorithm. }

\subsection{Model Preliminary}\label{sec:section_model_preliminary}
The Joint Semantic Transfer Network (JSTN) works under a semi-supervised setting. More specifically, it involves a source NI domain that is defined as follows: 
\begin{equation}\label{equ:source_ni_definition}
  \begin{split}
    & \mathcal{D}_{SN} = \{\mathcal{X}_{SN}, \mathcal{Y}_{SN}\} = \{(x_{SNi}, y_{SNi})\}, \\
    & x_{SNi} \in \mathbb{R}^{d_{SN}}, y_{SNi} \in [1, K], i \in [1, n_{SN}] \,,
  \end{split}
\end{equation}
where the source NI domain contains $n_{SN}$ instances with their corresponding label, each instance is a $d_{SN}$-dimensional vector, and each label is within a total of $K$ categories. Similarly, the small-scale source II domain is defined in a similar way as follows: 
\begin{equation}\label{equ:source_ii_definition}
  \begin{split}
    & \mathcal{D}_{SI} = \{\mathcal{X}_{SI}, \mathcal{Y}_{SI}\} = \{(x_{SIi}, y_{SIi})\}, \\
    & x_{SIi} \in \mathbb{R}^{d_{SI}}, y_{SIi} \in [1, K], i \in [1, n_{SI}], n_{SI} < n_{SN} \,.
  \end{split}
\end{equation}
Note that the amount of instances in the source II domain is smaller than the source NI domain due to data scarcity of IoT domains. Together, both the source NI domain $\mathcal{D}_{SN}$ and the source II domain $\mathcal{D}_{SI}$ form the source domain $\mathcal{D}_{S} = \{\mathcal{X}_{S}, \mathcal{Y}_{S}\} = \mathcal{D}_{SN} \cup \mathcal{D}_{SI}$, $n_S = n_{SN} + n_{SI}$. Under the semi-supervised setting, the target II domain is scarcely-labelled, and is defined as follows: 
\begin{equation}\label{equ:target_ii_definition}
  \begin{split}
    & \mathcal{D}_{TL} = \{\mathcal{X}_{TL}, \mathcal{Y}_{TL}\} = \{(x_{TLi}, y_{TLi})\}, \\
    & \mathcal{D}_{TU} = \{\mathcal{X}_{TU}\} = \{(x_{TUj})\}, \mathcal{D}_{T} = \mathcal{D}_{TL} \cup \mathcal{D}_{TU}\\
    & x_{TLi}, x_{TUj} \in \mathbb{R}^{d_{T}}, y_{TLi} \in [1, K], i \in [1, n_{TL}], j \in [1, n_{TU}] \\
    & n_{T} = n_{TL} + n_{TU}, n_{TL} \ll n_{TU}\,,
  \end{split}
\end{equation}
where only a small amount of target II \rw{data is labelled}, i.e., $n_{TL} \ll n_{TU}$. The source NI domain, source II domain and the target II domain \rw{present heterogeneities as} they come from distinct feature spaces, i.e., $d_{SN} \neq d_{SI} \neq d_{T}$. \rw{All notations used in this paper, and their corresponding interpretations, are presented in the Appendix to ease understanding. }

\subsection{\rw{JSTN Model Architecture}}\label{sec:section_model_architecture}
The architecture of the JSTN model is illustrated in Figure \ref{fig:figure_network_framework}. For each domain, a feature encoder $E$ is utilised to map the original feature into a shared common feature subspace with dimension $d_C$. The feature encoder $E$ is defined as follows: 
\begin{equation}
  \begin{split}
    & f(x_i) = \begin{cases}
      E_{SN}(x_i) & \text{if $x_i \in \mathcal{X}_{SN}$} \\
      E_{SI}(x_i) & \text{if $x_i \in \mathcal{X}_{SI}$} \\
      E_{T}(x_i) & \text{if $x_i \in \mathcal{X}_{T} = \mathcal{X}_{TL} \cup \mathcal{X}_{TU}$}
    \end{cases} \\
    & f(x_i) \in \mathbb{R}^{d_C}\,.
  \end{split}
\end{equation}
Instead of aligning heterogeneous domains into a common feature subspace via \rw{brute-force} and impair the feature discriminability, we apply a joint semantic transfer strategy to achieve a more fine-grained knowledge transfer. Specifically, the scenario semantic transfer partially masks heterogeneities between source NI and II domain by confusing the domain discriminator $D$ to produce domain-invariant common feature subspace. Meanwhile the categorical distribution knowledge is also transferred between source domains. Additionally, the weighted implicit semantic transfer is used to transfer the correlation relationships between category distributions so that the category distribution semantic will be preserved by the target and different categories will \rw{not be confounded mistakenly during transfer}. The knowledge from source domains are weighted \rw{based on their divergence with the target domain} to adaptively emphasise \rw{varied source importances} which reflect the degree of knowledge learning. Moreover, the hierarchical explicit semantic alignment is utilised to explicitly minimise the gap between instances of the same category from different domains via a global centroid-level alignment and a local representative-level alignment to increase discriminability. To fully explore the potentials of unlabelled target II domain instances while avoiding negative transfer caused by wrongly-assigned pseudo-labels, a pseudo-label refiner with an ensemble mechanism is used to leverage the geometric similarity information to enhance the pseudo-label accuracy. Finally, the labelled data will provide the supervision for training via a globally shared classifier $C$. The ultimate goal of the model is to use the trained shared classifier $C$ to work on the common feature subspace, so that the prediction accuracy of the unlabelled target II data is maximised.

\section{\rw{The JSTN Algorithm}}\label{sec:section_jstn_algorithm_details}

\rw{This section focuses on the detailed mechanisms of three JSTN constituting semantics, with their advantages explained in details. We then present the overall optimisation objective. }

\subsection{Scenario Semantic Transfer}\label{sec:section_scenario_semantic_transfer}

\textbf{\rw{1. The Auxiliary Source II Domain}}

\begin{figure}[t]
  \begin{center}
    \includegraphics[width=0.45\textwidth,keepaspectratio]{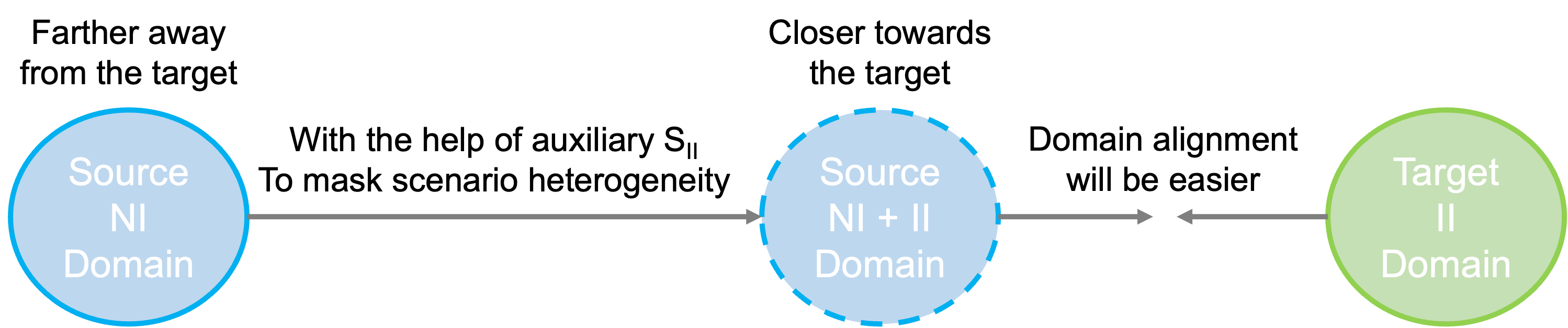}\\
    \caption{\rw{By introducing an auxiliary $S_{II}$ domain, it can partially mask the scenario heterogeneities between domains and hence ease the domain alignment process. }}
    \label{fig:figure_scenario_semantic_illustration}
    \vspace{-0.5cm}
  \end{center}
\end{figure}

The network intrusion data is rich in scale and intrusion knowledge, while the IoT intrusion data have rich IoT scenario characteristics. However, significant heterogeneities present between network and IoT intrusion domains \rw{as illustrated in Figure \ref{fig:figure_scenario_semantic_illustration}. }For instance, the network intrusion data are usually captured from servers in data centres, while IoT intrusion data comes from resource-constrained IoT infrastructures. \rw{Their diverse device types and working environment lead to heterogeneities such as different set of features, different feature dimensions, follow different distributions, etc. }On the other hand, although there are also heterogeneities between different II domains, however, the gap between II domains is smaller than the gap between NI and II domain. For example, although a fridge temperature monitor and a parcel GPS tracker have different functionalities, they usually work under similar network conditions compared with servers in top-tier data centres, they may utilise the same IoT network protocol that is different from servers, and hence the similarity between different II domains are higher than between NI and II domain. If we directly transfer the knowledge from NI domain to II domain via domain adaptation to facilitate intrusion detection, the huge domain gap between NI and II is still likely to hinder the effectiveness of knowledge transfer \rw{as in Figure \ref{fig:figure_scenario_semantic_illustration}}. Since \rw{it not only needs to} ensure fine-grained knowledge transfer, \rw{but also needs to} tackle significant divergences caused by different network protocol usages between domains, etc. 

However, if the source NI domain can be endowed with the IoT scenario semantic by using even a small amount of II data that is not from the target II domain due to target data scarcity, the gap between NI and II domain can be bridged more effectively, \rw{which can therefore benefit the source-target knowledge transfer performed later, as indicated in Figure \ref{fig:figure_scenario_semantic_illustration}. }Hence, to endow the NI domain with the characteristics of IoT domains, we use a small amount of II data from another IoT domain. 

\textbf{\rw{2. Scenario Semantic Transfer via Domain Discriminator}}

\rw{The JSTN performs scenario semantic transfer via a domain discriminator $D$. }When training the domain discriminator $D$, data instances from both the source NI and source II domain will be labelled as $1$, while the target II instances are labelled with $0$. By confusing the domain discriminator, the source NI and source II instances will be fused, so that the NI instances are equipped with IoT scenario semantic. \rw{When the domain discriminator $D$ is confused to distinguish the domain origin of instances, it promotes} a domain-invariant common feature subspace to be learned, which can benefit positive transfer. In the JSTN model, the domain discriminator is a neural network with a single layer that performs binary classification task with the loss defined as follows: 
\begin{equation}
  \begin{split}
    \mathcal{L}_{SSD} &= \frac{1}{n_{SN} + n_{SI}} \sum_{x_i \in \mathcal{X}_{SN} \cup \mathcal{X}_{SI}} \log(D(f(x_i)))\\
    & + \frac{1}{n_{TL} + n_{TU}} \sum_{x_j \in \mathcal{X}_{TL} \cup \mathcal{X}_{TU}} (1 - \log(D(f(x_j))))\,.
  \end{split}
\end{equation}
The domain encoders $E_{SN}$, $E_{SI}$ and $E_{T}$ will try to confuse the discriminator $D$ while the discriminator tries to stay unconfused. The common feature subspace yielded by domain encoders will become domain-invariant when this minimax game reaches an equilibrium. 

\textbf{\rw{3. Scenario Semantic Transfer via Distribution Matching}}

\begin{figure}[t]
  \begin{center}
    \includegraphics[width=0.45\textwidth,keepaspectratio]{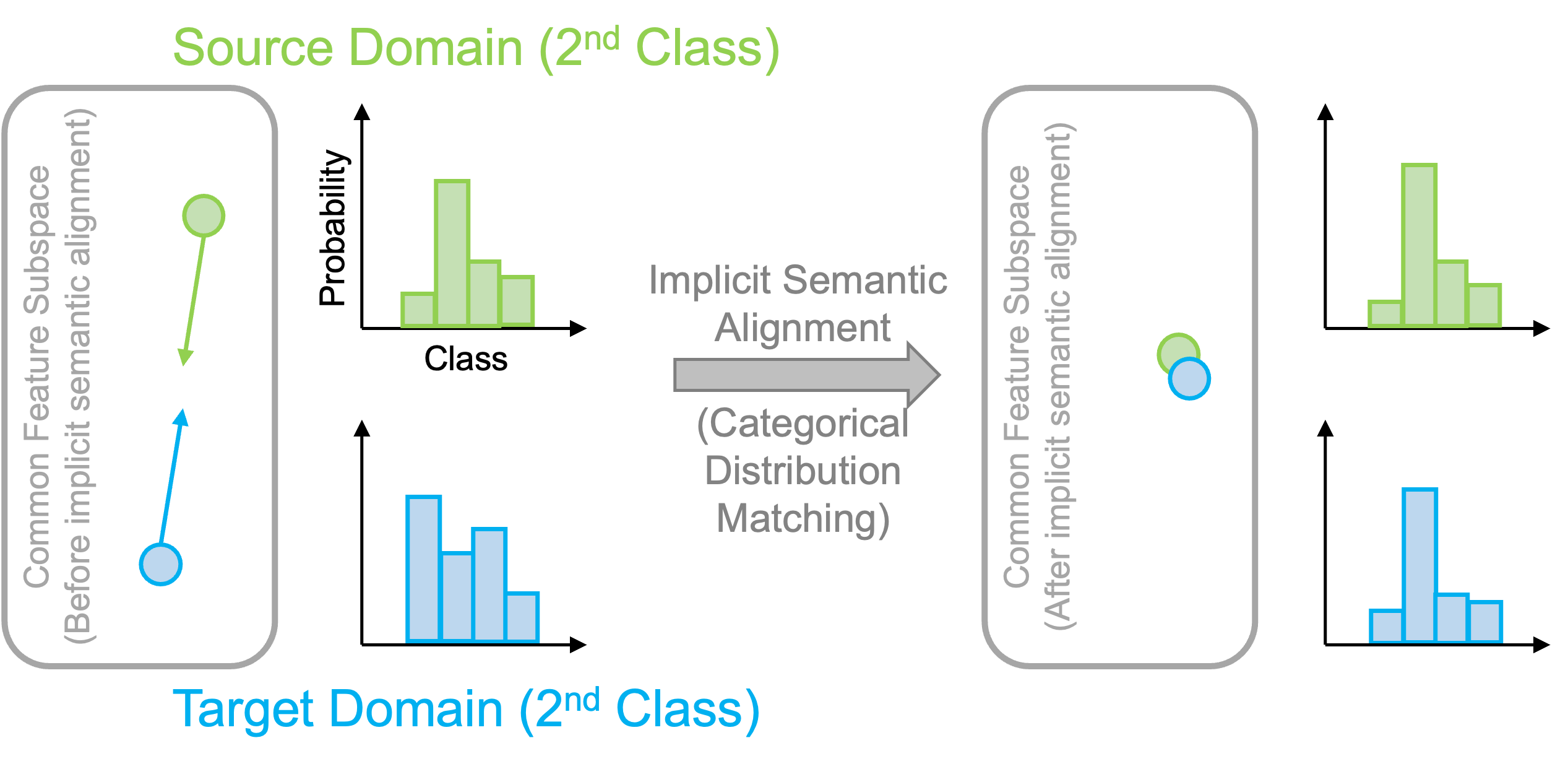}\\
    \caption{\rw{Before categorical distribution matching, the same category between domains does not align in the common feature subspace, and hence incur different categorical disbution correlations. After performing the categorical distribution matching, it forces the categorical distribution of each category to match between domains, which therefore enforces the category-wise alignment between domains in the common feature subspace. This figure uses instances in class 2 as an example. }}
    \label{fig:figure_implicit_semantic_alignment}
    \vspace{-0.5cm}
  \end{center}
\end{figure}

Besides, the source NI domain should also equip the source II domain with rich intrusion knowledge via the predicted categorical distribution knowledge transfer. Although during knowledge transfer, the source NI and II domain present heterogeneities, however, it is reasonable that instances from the same category should possess similar predicted categorical distribution, irrespective of which domain \rw{they come from}. Using objects \rw{as the example}, a PC monitor should be highly similar with other PC monitors, relatively similar with TV screens, and less likely to be similar with a bicycle or an orange, irrespective \rw{of its domain origin}. This preservation of categorical correlation applies for intrusion detection as well. Hence, transferring this distribution correlation knowledge will promote a more fine-grained feature alignment \rw{between domains in the common feature subspace}, \rw{as indicated in Figure \ref{fig:figure_implicit_semantic_alignment}. }Besides, it avoids mistaken category confounding, especially at the category boundaries in the \rw{learned feature space} that tend to be misinterpreted, \rw{since the categorical distribution can only be matched when the categories between domain align in the common feature subspace, as illustrated in Figure \ref{fig:figure_implicit_semantic_alignment}. Mathematically, the }source NI domain average probabilistic output of each category $k$ serves as the teacher to transfer the distribution correlation to the source II domain, and is defined as follows: 
\begin{equation}
  q^{(k)} = \frac{1}{|\mathcal{X}_{SN}^{(k)}|} \sum_{x_i \in \mathcal{X}_{SN}^{(k)}} softmax(\frac{C(f(x_i))}{T_1})\,,
\end{equation}
\rw{where $C(f(x_i))$ is the logit produced by the shared classifier, }$\mathcal{X}_{SN}^{(k)}$ represents the set of source NI instances belonging to the $k$\textsuperscript{th} category, $|\cdot|$ denotes the number of instances, and $T_1$ is a temperature hyperparameter that can smooth or sharp the categorical distribution during the semantic transfer. Similarly, the average probabilistic output of each category $k$ of the source II domain is defined as follows: 
\begin{equation}
  p^{(k)} = \frac{1}{|\mathcal{X}_{SI}^{(k)}|} \sum_{x_i \in \mathcal{X}_{SI}^{(k)}} softmax(\frac{C(f(x_i))}{T_1})\,.
\end{equation}
The distribution correlation knowledge is transferred from source NI domain to source II domain by minimising \rw{the divergence between $q^{(k)}$ and $p^{(k)}$ via the cross entropy loss defined as follows: }
\begin{equation}
  \mathcal{L}_{SSC} = - \frac{1}{k} \sum_{k = 1}^K {q^{(k)}}^{\top} \log(p^{(k)})\,,
\end{equation}

With the help of the scenario semantic, the network data can mimic the characteristics possessed by the IoT domain to some extent and increase its similarity with IoT domain, while the source II domain is also equipped with rich intrusion knowledge from the source NI domain. Both source domains will be drawn closer towards each other so that the significant heterogeneities will be partially masked, which will therefore ease the source-target knowledge transfer process. 

\subsection{Weighted Implicit Semantic Transfer}\label{sec:section_implicit_semantic_transfer}

\textbf{\rw{1. Implicit Semantic Transfer}}

Transferring the categorical distribution knowledge is useful not only between source domains, but also between source and target domains. The average probabilistic output of source instances belonging to category $k$ is treated as the teacher, or the ``soft label'' of category $k$, which is defined as follows: 
\begin{equation}
  q^{(k)}_{S*} = \frac{1}{|\mathcal{X}_{S*}^{(k)}|} \sum_{x_i \in \mathcal{X}_{S*}^{(k)}} softmax(\frac{C(f(x_i))}{T_2})\,,
\end{equation}
where $S* \in \{SN, SI\}$, \rw{$T_2$ is the smoothing temperature hyperparameter}. With the help of the soft label which contains the implicit semantic, we can let the probabilistic output of all labelled target instances $p_i$ to preserve the implicit semantic by minimising the soft loss defined as follows: 
\begin{equation}
  \begin{split}
    & p_i = softmax(C(f(x_i))), x_i \in \mathcal{X}_{TL} \\
    & \mathcal{L}_{sf}^{S*}(\mathcal{X}_{TL}, \mathcal{Y}_{TL}) = - \frac{1}{n_{TL}} \sum_{x_i \in \mathcal{X}_{TL}, y_i \in \mathcal{X}_{TL}} {q^{y_i}_{S*}}^{\top} \log(p_i)\,,
  \end{split}
\end{equation}
\rw{where $p_i$ is the categorical probabilistic output of the i\textsuperscript{th} labelled target instance, the soft loss $\mathcal{L}_{sf}^{S*}$ shortens the divergence of probabilistic outputs between domains. }Besides the soft label that is rich of implicit semantic, each labelled target II domain instance also has its corresponding label, i.e., the ``hard label''. The hard label will provide a supervised loss that is defined as follows: 
\begin{equation}
  \mathcal{L}_{hd}(\mathcal{X}_{TL}, \mathcal{Y}_{TL}) = \frac{1}{n_{TL}} \sum_{x_i \in \mathcal{X}_{TL}, y_i \in \mathcal{Y}_{TL}} \mathcal{L}_{ce}(C(f(x_i)), y_i)\,,
\end{equation}
\rw{where $\mathcal{L}_{ce}$ stands for cross entropy loss. }

\textbf{\rw{2. Divergence-based Weighting Scheme}}

Considering that the source NI and II domain may have different divergences towards the target II domain, which implicitly indicate their importance during implicit knowledge transfer, i.e., the \rw{degree of knowledge learning achieved} by the target II domain. The divergence between source domains and the target II domain \rw{$d_{<S*, TL>}$} are defined as follows:
\begin{equation}
  \begin{split}
    & \mu_{S*}^{(k)} = \frac{1}{|\mathcal{X}_{S*}|} \sum_{x_s \in \mathcal{X}_{S*}^{(k)}} f(x_s), \mu_{TL}^{(k)} = \frac{1}{|\mathcal{X}_{TL}|} \sum_{x_t \in \mathcal{X}_{TL}^{(k)}} f(x_t)\\
    & d_{<SN, TL>} = \frac{\sum_{k = 1}^K ||\mu_{SN}^{(k)} - \mu_{TL}^{(k)}||_2^2}{K}\\
    & d_{<SI, TL>} = \frac{\sum_{k = 1}^K ||\mu_{SI}^{(k)} - \mu_{TL}^{(k)}||_2^2}{K}\,,
  \end{split}
\end{equation}
\rw{where $\mu_{S*}^{(k)}$ stands for the class $k$ centroid of domain $S*$. }
Then, the weights \rw{$\omega_{<S*, TL>}$} for source domains during implicit knowledge transfer are defined as follows:
\begin{equation}
  \begin{split}
    & \omega_{<SN, TL>} = \frac{e^{d_{<SN, TL>}}}{e^{d_{<SN, TL>}} + 1} + 0.25\\
    & \omega_{<SI, TL>} = \frac{e^{d_{<SI, TL>}}}{e^{d_{<SI, TL>}} + 1} + 0.25\,.
  \end{split}
\end{equation}
The weight is controlled within the range between $0.75$ and $1.25$, \rw{so that source domains will neither completely loss its influence, nor have extremely heavy influence. The smaller the divergence is}, the smaller the weight is and vice versa. Therefore, if the source domain presents very little divergence with the target domain, then it indicates that the target domain already possesses the knowledge of that source domain, i.e., a relatively high degree of knowledge learning, and hence that source domain is suppressed using a smaller weight. Conversely, a large divergence between a source domain and the target domain indicates that the target domain is not yet fully equipped with the knowledge from that source domain, i.e., a relatively low degree of knowledge learning, and hence that source domain will be emphasised by a large weight. By utilising this weighting mechanism, it can dynamically adapt the relative importance of source domains during implicit knowledge transfer to maximise knowledge learning. 

Hence, the overall weighted implicit semantic loss $\mathcal{L}_{WIS}$ is defined as follows: 
\begin{equation}
  \begin{split}
    & \mathcal{L}_{WIS} = (1 - \alpha)\mathcal{L}_{hd}(\mathcal{X}_{TL}, \mathcal{Y}_{TL}) + \\
    & \frac{\alpha (\omega_{<SN, TL>} \mathcal{L}_{sf}^{SN}(\mathcal{X}_{TL}, \mathcal{Y}_{TL}) + \omega_{<SI, TL>} \mathcal{L}_{sf}^{SI}(\mathcal{X}_{TL}, \mathcal{Y}_{TL}))}{2}\,,
  \end{split}
\end{equation}
where hyperparameter $\alpha$ balances the influence between soft and hard losses. By optimising the weighted implicit semantic loss $\mathcal{L}_{WIS}$, the correlation within categorical distribution can be preserved by the target in the common feature subspace, and hence can prevent the negative transfer caused by confounded categories without enough discriminability. 

\subsection{Hierarchical Explicit Semantic Alignment}\label{sec:section_explicit_semantic_alignment}

\textbf{\rw{1. Pseudo-label Refiner (PLR)}}

The hierarchical explicit semantic alignment mechanism benefits the knowledge transfer by minimising the domain divergence from a distance perspective in a hierarchical manner. Considering that utilising the unlabelled target domain data during divergence minimisation would be helpful \cite{li2013learning,10.1145/3394171.3413995}, \rw{we perform the} pseudo-label assignment process before transferring the hierarchical explicit semantic. Several previous DA methods utilised pseudo-label for unlabelled target data, however, their pseudo-label assignment \rw{tended to be} inaccurate, which subsequently \rw{misled the model} training and caused negative transfer. To circumvent the negative effect caused by wrongly-assigned pseudo-label, we utilise a pseudo-label refiner (PLR) based on the ensemble paradigm to improve the pseudo-label assignment accuracy. For each unlabelled target II domain instance $x_{i} \in \mathcal{X}_{TU}$, the shared classifier will yield a prediction, which is treated as the neural network label, denoted as $y_{i}^{<NN>}$. Various previous DA efforts directly utilised the predicted label as pseudo-label assignment, however, these pseudo-labels are error-prone, especially during initial training stage. Therefore, we also take the intrinsic geometric knowledge into account. For both the source data and the labelled target data, we calculate the centroid of instances for each class $\mu^{(k)}$, which is defined as follows: 
\begin{equation}
  \begin{split}
    \mu^{(k)} &= \frac{1}{|\mathcal{X}_{SN} \cup \mathcal{X}_{SI} \cup \mathcal{X}_{TL}|} \\
    & (\sum_{x_l \in \mathcal{X}_{SN}^{(k)}} f(x_l) + \sum_{x_m \in \mathcal{X}_{SI}^{(k)}} f(x_m) + \sum_{x_n \in \mathcal{X}_{TL}^{(k)}} f(x_n))\,,
  \end{split}
\end{equation}
where $\mathcal{X}_{SN}^{(k)}$ means class $k$ $\mathcal{X}_{SN}$ instances. After obtaining the centroid of labelled instances for each category, we can assign each unlabelled target instance to the category whose centroid has the highest Cosine similarity with that unlabelled target instance, namely the geometric similarity-based (GS) label \rw{$y_{i}^{<GS>}$}. The GS label is decided as follows: 
\begin{equation}
  y_{i}^{<GS>} = \argmax_k \{CS(f(x_i), \mu^{(k)})\}, x_i \in \mathcal{X}_{TU}\,,
\end{equation}
where $CS()$ is the Cosine similarity. Considering that when the neural network label and the geometric similarity label reach a consensus, it gives the pseudo-label a stronger confidence to be correct since it is more unlikely for both the trained classifier and the intrinsic geometric property to reach the same wrong assignment simultaneously. Hence, the pseudo-label refiner forms a refinement mechanism. It will only assign a pseudo-label to the unlabelled target instance if an agreement is reach, or otherwise that unlabelled target instance will not have a pseudo-label assignment and will not be utilised during hierarchical explicit semantic transfer to circumvent error cumulation. We denote the assigned pseudo-label as $\hat{y_i}$ and denote its corresponding feature vector as $\hat{x_i}$. Hence, the target instance set will be updated as follows: 
\begin{equation}
  \begin{split}
    & \mathcal{X}_{T} = \mathcal{X}_{TL} \cup \{\hat{x_i}\}, \mathcal{Y}_{T} = \mathcal{Y}_{TL} \cup \{\hat{y_i}\}, \\
    & i \in [1, n_{TU}], y_{i}^{<NN>} = y_{i}^{<GS>}\,.
  \end{split}
\end{equation}
Initially, the model is not stable enough, hence only a few pseudo-label will be assigned and a majority of \rw{unconfident pseudo-label assignment} will be filtered out to prevent error cumulation. As the training progresses, assignment agreements will be reached for more unlabelled target instances, which will let them to participate in the hierarchical explicit semantic transfer. Eventually, at later training stage, a majority of unlabelled instances will be assigned with a consistent pseudo-label, which can make the unlabelled target instances be explored as much as possible. Hence, the pseudo-label refiner can filter out pseudo-label assignments that are possibly wrong to prevent negative transfer, it forms an automatic pseudo-label assignment process without requiring human experience or manually-assigned thresholds. 

\textbf{\rw{2. Hierarchical Explicit Semantic Transfer - Global Centroid Level}}

With the help of the PLR, we can perform the hierarchical explicit semantic transfer from two levels. Firstly, a global level triplet centroid alignment is performed to align category-wise centroids. Specifically, we can calculate the category-wise centroid for the source domain \rw{$\mu_{S}^{(k)}$}, target domain \rw{$\mu_{T}^{(k)}$}, and the combination of source and target domain \rw{$\mu_{ST}^{(k)}$}, which are defined as follows: 
\begin{equation}
  \begin{split}
    \mu_{S}^{(k)} &= \frac{1}{|\mathcal{X}_{SN}^{(k)} \cup \mathcal{X}_{SI}^{(k)}|} \sum_{x_i \in \mathcal{X}_{SN}^{(k)} \cup \mathcal{X}_{SI}^{(k)}} f(x_i) \\
    \mu_{T}^{(k)} &= \frac{1}{|\mathcal{X}_{T}^{(k)}|} \sum_{x_j \in \mathcal{X}_{T}^{(k)}} f(x_j) \\
    \mu_{ST}^{(k)} &= \frac{1}{|\mathcal{X}_{SN}^{(k)} \cup \mathcal{X}_{SI}^{(k)} \cup \mathcal{X}_{T}^{(k)}|} \\
    & (\sum_{x_i \in \mathcal{X}_{SN}^{(k)} \cup \mathcal{X}_{SI}^{(k)}} f(x_i) + \sum_{x_j \in \mathcal{X}_{T}^{(k)}} f(x_j))\,.
  \end{split}
\end{equation}
\rw{Then, we explicitly} learn a more robust and discriminative feature representation by minimising the intra-category divergence, i.e., minimising the \rw{$L_2$-distances between} each centroid, which is defined as follows:
\begin{equation}
  \mathcal{L}_{ESC} = \sum_{k=1}^{K} (||\mu_{S}^{(k)} - \mu_{T}^{(k)}||_2^2 + ||\mu_{S}^{(k)} - \mu_{ST}^{(k)}||_2^2 + ||\mu_{T}^{(k)} - \mu_{ST}^{(k)}||_2^2)\,.
\end{equation}

\begin{figure}[t]
  \begin{center}
    \includegraphics[width=0.42\textwidth,keepaspectratio]{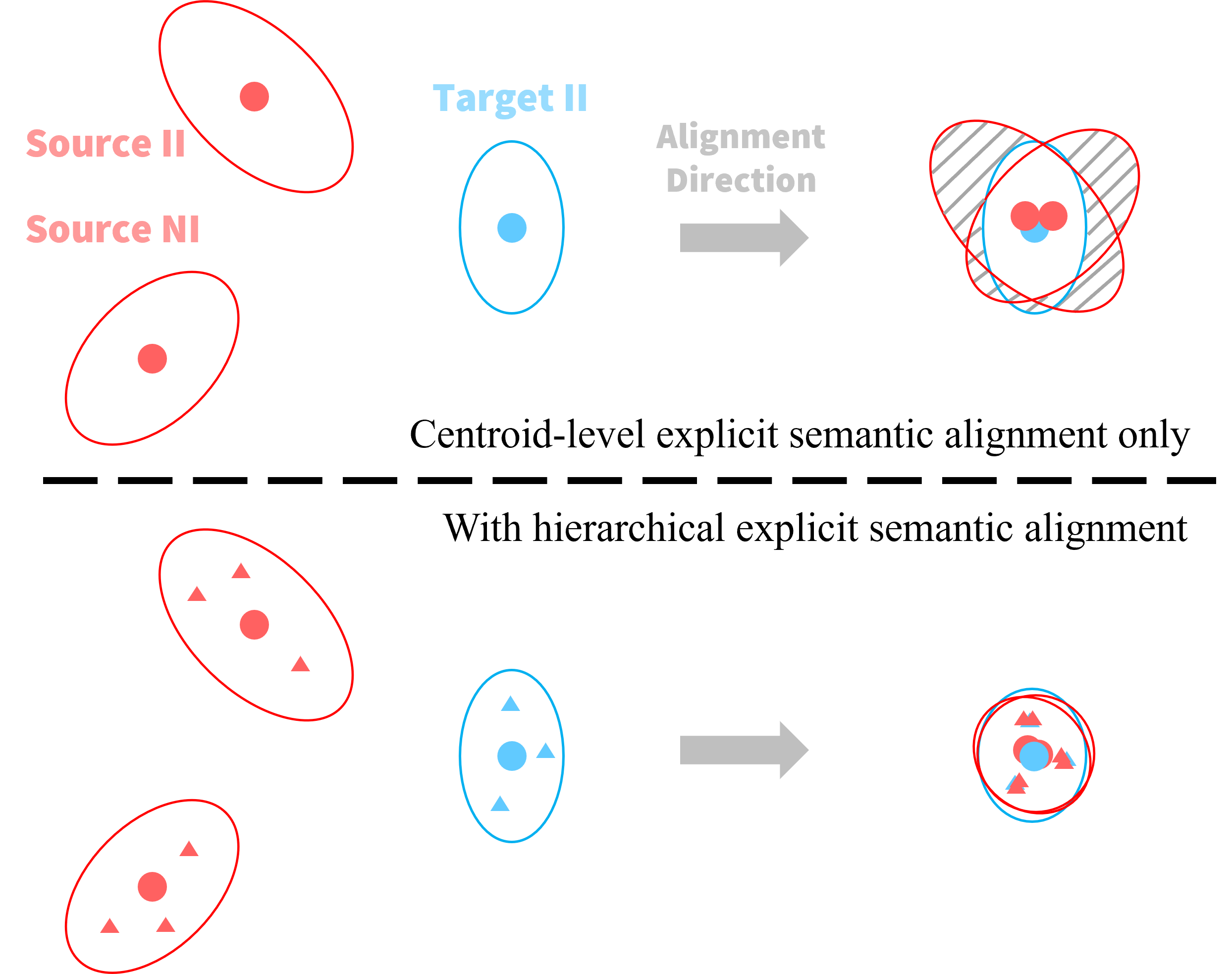}\\
    \caption{Illustration of alignment effect of ($\uparrow$) only apply centroid-level explicit semantic alignment and ($\downarrow$) apply the hierarchical explicit semantic alignment. The dot and rectangle indicate class centroid and class representative points, respectively. For simplified illustration, only a single class is plotted. }
    \label{fig:figure_hierarchical_explicit_semantic_alignment}
    \vspace{-0.5cm}
  \end{center}
\end{figure}

\textbf{\rw{3. Hierarchical Explicit Semantic Transfer - Local Representative Level}}

However, \rw{only performing the} global-level centroid alignment is not enough to achieve fine-grained semantic consistency. As shown in the upper Figure \ref{fig:figure_hierarchical_explicit_semantic_alignment}, even though the centroids are alignment, the features can still lack concentration as indicated by the grey shaded area, which hurts the semantic transfer. This is due to that centroids only represent the category at the global level, which lack thorough coverage of the whole category in a fine-grained manner. Therefore, the explicit semantic alignment is also performed from a local perspective to achieve a more fine-grained category coverage. For each category $k$ in both the source domain and target domain ($\mathcal{D}_{T} = (\mathcal{X}_{T}, \mathcal{Y}_{T})$), $R$ representatives are selected via Kmeans$++$ clustering, denoted as $r^k_{S}(i)$ and $r^k_{T}(i), k \in [1, K], i \in [1, R]$, respectively. Then, we calculate the pairwise distances between source and target representatives for each category as follows: 
\begin{equation}
  \mathcal{L}_{ESR} = \sum_{k=1}^{K} \frac{\sum_{i=1}^{R} \sum_{j=1}^{R} ||r^k_{S}(i) - r^k_{T}(j)||_2^2}{K \times |r^k_{S}| \times |r^k_{T}|}
\end{equation}
Unlike performing the pairwise divergence minimisation for all source and target instances as in \cite{li2021semantic_iccv}, the hierarchical explicit semantic works on category-wise representatives, which avoids the severe computational burden without hurting the alignment effectiveness. By explicitly minimising the intra-category divergence \rw{from the global-level} centroid perspective and the local-level representative perspective, each domain in the common feature subspace will be more semantically consistent in a concentrated manner as indicated in the lower part of Figure \ref{fig:figure_hierarchical_explicit_semantic_alignment}. 

\subsection{Overall Optimisation Objective}\label{sec:section_overall_optimisation_objective}
Finally, the ground truth labels of source domains and the predicted output yielded by the shared classifier will produce a supervision loss \rw{$\mathcal{L}_{sup}$} as follows:
\begin{equation}
  \mathcal{L}_{sup}(\mathcal{X}_{S}, \mathcal{Y}_{S}) = \frac{1}{n_{S}} \sum_{x_i \in \mathcal{X}_{S}, y_i \in \mathcal{Y}_{S}} \mathcal{L}_{ce}(C(f(x_i)), y_i)\,,
\end{equation}
while the supervision loss of the labelled target domain has been treated as hard label loss in the implicit semantic transfer as previously mentioned. Overall, the optimisation objective of the JSTN model is as follows: 
\begin{equation}
  \begin{split}
    \min_{C, E_{SN}, E_{SI}, E_{T}} \max_{D} &\{\mathcal{L}_{sup} + \mathcal{L}_{WIS} + \beta \mathcal{L}_{ESC} + \\
    & \lambda \mathcal{L}_{ESR} + \gamma \mathcal{L}_{SSD} + \eta \mathcal{L}_{SSC}\}\,,
  \end{split}
\end{equation}
where $\beta$, $\lambda$, $\gamma$ and $\eta$ are hyperparameters that control the influence of loss components during optimisation. \rw{Inspired by \cite{ganin2015unsupervised}, we apply the Gradient Reversal Layer (GRL) on the discriminator to train the entire JSTN network in an end-to-end manner using Adam gradient descent. }By optimising the overall objective, the scenario semantic fuses domains by confusing domain discriminator and meanwhile transfer knowledge between source domains. Hence, it forms a domain-invariant feature subspace so that the heterogeneities between source-source domains and source-target domains will be minimised. The weighted implicit semantic increases generalisability through preserving the implicit categorical distribution knowledge, the knowledge from different source domains are weighted \rw{based on their relative} divergence with the target domain to indicate source importance which reflects the degree of knowledge learning. Meanwhile, the hierarchical explicit semantic \rw{learns} a robust and semantically consistent common feature subspace with compactness and concentration from global centroid perspective and local representative perspective so that the intra-category divergence will be shortened. By jointly leveraging these semantics, the knowledge transfer effectiveness of the model will be enhanced. Upon the above minimax game reaches an equilibrium, the training of feature encoders for each domain, the shared classifier $C$ and the domain discriminator $D$ concludes. 

\section{Experiment}\label{sec:section_experiment}

\begin{table*}[!ht]
  \caption{Intrusion detection accuracy of $10$ methods on $10$ tasks. }
  \vspace{-0.3cm}
  \centering
  {\renewcommand{\arraystretch}{1.3}
  \begin{tabular}{c|cccccccccc|c}
  \Xhline{2\arrayrulewidth}
  \multicolumn{1}{p{1.5cm}}{\centering $S_N + S_I$\\$\rightarrow T_I$} & \multicolumn{1}{p{1cm}}{\centering $C + M$\\$\rightarrow B$} & \multicolumn{1}{p{1cm}}{\centering $C + W$\\$\rightarrow B$} & \multicolumn{1}{p{1cm}}{\centering $C + B$\\$\rightarrow G$} & \multicolumn{1}{p{1cm}}{\centering $N + G$\\$\rightarrow B$} & \multicolumn{1}{p{1cm}}{\centering $N + F$\\$\rightarrow B$} & \multicolumn{1}{p{1cm}}{\centering $N + B$\\$\rightarrow W$} & \multicolumn{1}{p{1cm}}{\centering $N + M$\\$\rightarrow B$} & \multicolumn{1}{p{1cm}}{\centering $K + M$\\$\rightarrow B$} & \multicolumn{1}{p{1cm}}{\centering $K + W$\\$\rightarrow B$} & \multicolumn{1}{p{1cm}}{\centering $K + M$\\$\rightarrow G$} & \multicolumn{1}{p{0.6cm}}{\centering Avg} \\ \hline
  SB-RF & $48.84$ & $46.76$ & $73.40$ & $47.36$ & $64.87$ & $82.30$ & $50.50$ & $48.81$ & $46.65$ & $82.30$ & $59.18$ \\
  SB-SVM & $52.56$ & $50.96$ & $73.60$ & $51.78$ & $67.84$ & $82.40$ & $55.00$ & $52.55$ & $50.28$ & $82.33$ & $61.93$ \\
  SB-NN & $53.01$ & $51.58$ & $73.68$ & $47.51$ & $65.38$ & $82.36$ & $56.62$ & $52.53$ & $51.70$ & $82.35$ & $61.67$ \\
  SB-TNT & $48.22$ & $25.57$ & $73.50$ & $25.93$ & $35.25$ & $82.24$ & $49.50$ & $48.22$ & $24.94$ & $82.25$ & $49.56$ \\
  SB-DDAC & $60.71$ & $53.62$ & $72.84$ & $52.88$ & $72.32$ & $85.60$ & $63.35$ & $60.71$ & $53.05$ & $82.35$ & $65.74$ \\
  SB-DDAS & $58.98$ & $50.84$ & $73.68$ & $52.21$ & $70.21$ & $85.54$ & $61.50$ & $58.98$ & $50.89$ & $82.34$ & $64.52$ \\
  SB-STN & $62.29$ & $54.92$ & $76.83$ & $54.99$ & $71.83$ & $87.00$ & $63.95$ & $62.79$ & $54.44$ & $85.57$ & $67.46$ \\
  STN & $61.77$ & $54.45$ & $76.18$ & $55.20$ & $73.69$ & $86.91$ & $63.79$ & $60.08$ & $54.69$ & $85.15$ & $67.19$ \\
  CWAN & $59.98$ & $52.40$ & $75.84$ & $53.37$ & $72.60$ & $85.17$ & $59.69$ & $60.29$ & $52.99$ & $85.53$ & $65.79$ \\ \hline
  \textbf{JSTN (Ours)} & $\textbf{67.27}$ & $\textbf{69.69}$ & $\textbf{77.08}$ & $\textbf{70.20}$ & $\textbf{86.94}$ & $\textbf{87.78}$ & $\textbf{69.31}$ & $\textbf{66.95}$ & $\textbf{69.73}$ & $\textbf{87.10}$ & $\textbf{75.21}$ \\ \Xhline{2\arrayrulewidth}
  \end{tabular}}
  \label{tab:giant_performance_table}
  \vspace{-0.3cm}
\end{table*}

\subsection{Dataset and Setup}\label{sec:section_dataset_and_setup}

\textbf{Network Intrusion Dataset: NSL-KDD} The NSL-KDD (K) \cite{tavallaee2009nslkdd} network intrusion dataset was released in 2009. It improves the outdated KDD99 dataset \cite{hettich1999uci} to reflect modern network attack characteristics. It contains benign traffic with $4$ malicious attack categories, such as denial of service (DoS), probing attack, etc. It does not have redundant or duplicate records, the quality of data is significantly improved. Follow \cite{anthi2019smart_home_iot}, we use $20\%$ of the dataset, which is a reasonable and affordable amount. Each record is represented using $41$ features. We follow Harb et al., \cite{harb2011selecting} to choose $31$ most informative features as the feature space. 

\textbf{Network Intrusion Dataset: UNSW-NB15} The UNSW-NB15 (N) network intrusion dataset \cite{moustafa2015unsw} was created in 2015 using the IXIA PerfectStorm tool. The dataset also aims to tackle the limitations such as redundant records or missing values of previous IDS datasets, especially under a modern low footprint environment. The dataset contains benign network behaviours plus $9$ attack categories, such as DoS attack, reconnaissance attack, etc. The dataset contains $257,646$ records, follow previous work \cite{alkadi2020deep}, we utilise $6000$ entries during the model training and evaluation. Each record is represented using $49$ features. The preprocessing steps include removing $4$ features that have value $0$ for nearly all records. 

\textbf{Network Intrusion Dataset: CICIDS2017} The CICIDS2017 (C) \cite{sharafaldin2018toward} network intrusion dataset was released in 2017. It is one of the most up-to-date network traffic datasets. The data is collected using CICFlowMeter. The dataset has benign and $7$ common intrusion attack types to reflect the current trend. The attack types including DoS, distributed DoS (DDoS), Brute Force attack, etc. Portion of the dataset ($20\%$) \cite{stiawan2020cicids} has been provided in CSV format for ML training, represented using $77$ features. We perform preprocessing including data deduplication, and converting categorical attributes to numerical entries. We follow \cite{anthi2019smart_home_iot} to utilise $20000$ entries of network traffic, a reasonable amount to train an effective IDS. Guided by the information gain-based feature selection work of Stiawan \cite{stiawan2020cicids} et al., we use the features with top $40$ information gain, which can effectively filter out information-scarce features and improve training efficiency. 

\textbf{IoT Intrusion Dataset: UNSW-BOTIOT} The UNSW-BOTIOT (B) dataset \cite{koroniotis2019towards} was also created in 2017 with a specific focus on realistic IoT intrusion scenarios. It applies $5$ IoT scenarios in the testbed, including a weather station, a smart fridge, a smart thermostat associated with in-house air-conditioning, etc. The testbed also utilises the MQTT protocol, a lightweight communication protocol commonly used between IoT devices. Hence, the dataset fills the void of lacking specific consideration for IoT scenarios. It contains $3$ common IoT attack categories, such as DoS attack, information theft, etc. Following \cite{alkadi2020deep}, we utilise $7500$ records during the model training and evaluation. Note that when used as source II domain, the amount of data is $1/6 - 1/2$ of the amount of source NI data to reflect the reality that IoT intrusion data is scarcer than network intrusion data. Besides, under the semi-supervised setting, we follow \cite{10.1145/3394171.3413995,ning2021malware,yao2019heterogeneous} to vary the $n_{TL}:n_{TU}$ ratio among $1:2$, $1:10$ and $1:50$, i.e., the amount of unlabelled target II data is much higher than the amount of labelled target II data. Each record in the dataset is represented using $46$ features. Following the official suggestion \cite{koroniotis2019towards}, we utilise the top $10$ most informative features to represent each record. 

\textbf{IoT Intrusion Dataset: UNSW-TONIOT} The UNSW-TONIOT dataset \cite{booij2021ton_iot} was another popular IoT intrusion dataset \cite{abdelmoumin2021performance} released in 2021. It contains IoT intrusion data that comply with the protocols, standards and technologies commonly used by current IoT devices. It further extends the number of IoT devices used in the testbed and the diversity of attacks being considered. The testbed operates $7$ IoT sensors such as weather monitor, smart fridge monitor, Modbus sensor, GPS tracker, etc., and the dataset covers $9$ kinds of threats, including scanning attack, DoS attack, etc. To reflect the heterogeneities of IoT devices, each IoT device in the dataset has its own set of features, e.g., the smart garage door will record the door state, and whether the door receives a control signal from the phone app, while the GPS tracker will record the latitude and longitude of the object it attaches on. We select $4$ representative IoT devices (weather monitor, modbus sensor, GPS tracker and fridge monitor) and in total utilise $21900$ records from the dataset, which account for around $10\%$ of the data and is reasonable for model training and evaluation \cite{alkadi2020deep,qiu2020adversarial}. Following the setting of UNSW-BOTIOT dataset, when being used as the source II domain, the amount of data is around $1/5 - 1/2$ compared with the source NI domain data, and the $n_{TL}:n_{TU}$ ratio is varied among $1:5$, $1:10$ and $1:50$. The UNSW-TONIOT dataset is abbreviated as ``W'', ``M'', ``G'' and ``F'' for TONIOT weather monitor, modbus sensor, GPS tracker and fridge monitor, respectively. 

\textbf{Shared Intrusion} To perform knowledge transfer to facilitate the intrusion detection for the target II domain, $5$ shared categories are picked out from the aforementioned datasets, namely benign class, DoS attack, DDoS attack, reconnaissance attack and password attack. These shared common categories are representatives of the majority of modern intrusions faced by networks and IoT devices, they account for $99.85\%$, $99.96\%$, $54.2\%$, $100\%$ and $77.03\%$ amounts of records in the CICIDS2017, NSL-KDD, UNSW-NB15, UNSW-BOTIOT and UNSW-TONIOT dataset, respectively. Therefore, after transferring the knowledge, most modern intrusion attacks faced by the IoT domain can be detected. 

\begin{table*}[!ht]
  \caption{Intrusion detection accuracy under varied $n_{TL}:n_{TU}$ ratios. }
  \vspace{-0.3cm}
  \centering
  {\renewcommand{\arraystretch}{1.3}
  \begin{tabular}{c|ccccccccc}
    \Xhline{2\arrayrulewidth}
    \multicolumn{1}{p{1.8cm}|}{\centering $S_N + S_I \rightarrow T_I$} & \multicolumn{3}{p{3cm}}{\centering $N + M \rightarrow G$} & \multicolumn{3}{p{3cm}}{\centering $K + B \rightarrow W$} & \multicolumn{3}{p{3cm}}{\centering $N + G \rightarrow W$} \\ \hline
    $n_{LT} : n_{UT}$ & $1:5$ & $1:10$ & $1:50$ & $1:5$ & $1:10$ & $1:50$ & $1:5$ & $1:10$ & $1:50$ \\ \hline
    SB-RF & $93.33$ & $93.32$ & $93.28$ & $73.68$ & $73.65$ & $73.61$ & $82.36$ & $82.35$ & $82.33$ \\
    SB-SVM & $93.32$ & $93.29$ & $93.26$ & $75.74$ & $75.36$ & $73.64$ & $85.15$ & $85.07$ & $82.34$ \\
    SB-NN & $93.33$ & $93.32$ & $93.30$ & $73.69$ & $73.62$ & $73.51$ & $84.13$ & $84.07$ & $83.78$ \\
    SB-TNT & $93.30$ & $93.28$ & $93.22$ & $73.54$ & $73.53$ & $73.52$ & $82.25$ & $82.24$ & $82.22$ \\
    SB-DDAC & $93.33$ & $93.33$ & $93.31$ & $78.92$ & $78.90$ & $78.65$ & $85.60$ & $85.58$ & $85.50$ \\
    SB-DDAS & $93.33$ & $93.32$ & $93.30$ & $75.81$ & $75.70$ & $75.63$ & $85.71$ & $85.52$ & $85.38$ \\
    SB-STN & $93.48$ & $93.33$ & $93.19$ & $79.55$ & $79.50$ & $76.85$ & $86.65$ & $86.53$ & $85.48$ \\
    STN & $92.99$ & $92.94$ & $92.39$ & $79.52$ & $78.76$ & $77.92$ & $86.53$ & $86.52$ & $86.37$ \\
    CWAN & $93.50$ & $93.33$ & $93.00$ & $78.46$ & $78.42$ & $78.02$ & $85.80$ & $85.38$ & $84.63$ \\ \hline
    \textbf{JSTN (Ours)} & $\textbf{94.46}$ & $\textbf{94.01}$ & $\textbf{93.96}$ & $\textbf{80.50}$ & $\textbf{79.95}$ & $\textbf{78.84}$ & $\textbf{87.34}$ & $\textbf{87.26}$ & $\textbf{87.19}$ \\ \Xhline{2\arrayrulewidth}
    \end{tabular}}
  \label{tab:ratio_performance_table1}
  \vspace{-0.2cm}
\end{table*}

\begin{table*}[!ht]
  \caption{(Continued from Table \ref{tab:ratio_performance_table1}) Intrusion detection accuracy under varied $n_{TL}:n_{TU}$ ratios. }
  \vspace{-0.3cm}
  \centering
  {\renewcommand{\arraystretch}{1.3}
  \begin{tabular}{c|ccccccccc|cc}
    \Xhline{2\arrayrulewidth}
    \multicolumn{1}{p{1.8cm}|}{\centering $S_N + S_I \rightarrow T_I$} & \multicolumn{3}{p{3cm}}{\centering $C + W \rightarrow B$} & \multicolumn{3}{p{3cm}}{\centering $N + M \rightarrow B$} & \multicolumn{3}{p{3cm}|}{\centering $N + W \rightarrow B$} & \multirow{2}{*}{\centering Overall Avg} & \multirow{2}{*}{\centering $1:50$ Case Avg} \\ \cline{1-10}
    $n_{LT} : n_{UT}$ & $1:2$ & $1:10$ & $1:50$ & $1:2$ & $1:10$ & $1:50$ & $1:2$ & $1:10$ & $1:50$ & & \\ \hline
    SB-RF & $46.76$ & $46.70$ & $46.53$ & $50.50$ & $50.22$ & $50.08$ & $47.38$ & $47.37$ & $47.35$ & $65.60$ & $65.53$ \\
    SB-SVM & $50.96$ & $49.49$ & $48.51$ & $55.03$ & $54.10$ & $53.49$ & $51.47$ & $50.04$ & $48.98$ & $67.74$ & $66.70$ \\
    SB-NN & $51.58$ & $48.67$ & $48.50$ & $56.62$ & $53.59$ & $53.20$ & $51.50$ & $50.92$ & $50.67$ & $67.67$ & $67.16$ \\
    SB-TNT & $25.57$ & $25.45$ & $24.99$ & $49.50$ & $49.33$ & $49.00$ & $25.93$ & $25.78$ & $25.60$ & $58.24$ & $58.09$ \\
    SB-DDAC & $53.62$ & $52.55$ & $51.83$ & $63.35$ & $62.05$ & $60.80$ & $52.95$ & $52.79$ & $52.75$ & $70.88$ & $70.47$ \\
    SB-DDAS & $51.84$ & $51.60$ & $49.80$ & $61.50$ & $59.45$ & $58.70$ & $51.97$ & $51.90$ & $51.30$ & $69.54$ & $69.02$ \\
    SB-STN & $54.92$ & $54.47$ & $52.74$ & $63.95$ & $62.19$ & $62.07$ & $54.60$ & $54.28$ & $53.01$ & $71.49$ & $70.56$ \\
    STN & $54.29$ & $53.06$ & $51.78$ & $61.98$ & $60.91$ & $55.94$ & $54.08$ & $53.24$ & $52.61$ & $70.66$ & $69.50$ \\
    CWAN & $52.40$ & $51.87$ & $51.77$ & $59.69$ & $58.32$ & $56.28$ & $53.03$ & $52.12$ & $52.10$ & $69.90$ & $69.30$ \\ \hline
    \textbf{JSTN (Ours)} & $\textbf{69.69}$ & $\textbf{66.49}$ & $\textbf{61.96}$ & $\textbf{69.31}$ & $\textbf{67.71}$ & $\textbf{67.65}$ & $\textbf{69.93}$ & $\textbf{67.24}$ & $\textbf{67.15}$ & $\textbf{77.26}$ & $\textbf{76.13}$ \\ \Xhline{2\arrayrulewidth}
    \end{tabular}}
  \label{tab:ratio_performance_table2}
  \vspace{-0.3cm}
\end{table*}

\textbf{Implementation Details} We implement the JSTN model using the PyTorch \cite{paszke2019pytorch} DL framework, and deploy the experiment on a server equipped with Intel Core i9-9900K CPU and Nvidia Tesla V100 GPU. All feature encoders are two-layer fully-connected neural networks with LeakyReLU \cite{maas2013rectifier} as the activation function following \cite{10.1145/3394171.3413995,yao2019heterogeneous}. Both the shared classifier $C$ and the domain discriminator $D$ are single layer neural networks. For hyperparameter settings, we empirically set $\alpha = 0.1$, $\beta = 0.004$, $\lambda = 0.001$, $\gamma = 0.1$, $\eta = 0.001$, $T_1 = 10$, $T_2 = 5$, and the dimension of the domain-invariant common feature subspace $d_{C}$ is set to $3$. We also verify the parameter sensitivity in Section \ref{sec:section_parameter_sensitivity_analysis} to indicate the JSTN can perform stably and robustly under varied parameter settings. \rw{Follow \cite{10.1145/3394171.3413995}, we optimise the JSTN model using Adam gradient descent optimiser and set the number of epochs to $1000$. }Following \cite{alkadi2020deep,li2018ai_accuracy}, we mainly use accuracy on the unlabelled target II data as the evaluation metrics, as well as category-weighted precision (P), recall (R) and F1-score (F) \cite{ferrag2020deep} to evaluate the performance. \rw{We define true positive $TP^{(k)}$ to be the number of unlabelled target II instances which belong to intrusion category $k$ and are correctly predicted as intrusion category $k$, similar for true negative $TN^{(k)}$, false positive $FP^{(k)}$ and false negative $FN^{(k)}$. Hence, the category-weighted precision, recall and F1-score are mathematically defined as follows: }

\rw{\begin{equation}
  \begin{split}
    Precision & = \sum_{k=1}^{K} \frac{|\mathcal{X}_{TU}^{(k)}|}{n_{TU}} \cdot \frac{TP^{(k)}}{TP^{(k)} + FP^{(k)}}\\
    & = \sum_{k=1}^{K} \frac{|\mathcal{X}_{TU}^{(k)}|}{n_{TU}} \cdot Precision^{(k)}\,,
  \end{split}
\end{equation}}

\rw{\begin{equation}
  \begin{split}
    Recall & = \sum_{k=1}^{K} \frac{|\mathcal{X}_{TU}^{(k)}|}{n_{TU}} \cdot \frac{TP^{(k)}}{TP^{(k)} + FN^{(k)}}\\
    & = \sum_{k=1}^{K} \frac{|\mathcal{X}_{TU}^{(k)}|}{n_{TU}} \cdot Recall^{(k)}\,,
  \end{split}
\end{equation}}

\rw{\begin{equation}
  \begin{split}
    & F1 = \sum_{k=1}^{K} \frac{|\mathcal{X}_{TU}^{(k)}|}{n_{TU}} \cdot \frac{2 \cdot Precision^{(k)} \cdot Recall^{(k)}}{Precision^{(k)} + Recall^{(k)}}\,.
  \end{split}
\end{equation}}

\textbf{Baseline Methods} We utilise $6$ state-of-the-art HDA methods as our comparing methods, including the CWAN \cite{yao2021multisource}, which is capable to transfer knowledge from two source domains to the target domain, the double-source STN \cite{yao2019heterogeneous}, as well as the single-source STN, TNT \cite{chen2016transfer}, DDAC and DDAS \cite{yao2020discriminative}, which can transfer knowledge from a single source domain to the target domain. For these $4$ single-source methods, we perform the $S_{NI} \rightarrow T_{II}$ and $S_{II} \rightarrow T_{II}$ transfer, and use the higher result as their final evaluation result, which is denoted as single-best (SB) during experiment. Besides, $3$ machine learning techniques are also utilised, including a two-layer neural network, support vector machine and random forest. These three machine learning learners are trained using the labelled target II data, and then perform ID on the unlabelled target II domain. They are denoted as NN, SVM and RF, respectively. 

\begin{table}[!t]
  \caption{Intrusion detection precision, recall and F1-score of DA-based methods. }
  \vspace{-0.3cm}
  \centering
  {\renewcommand{\arraystretch}{1.3}
  \begin{tabular}{c|cccccc}
    \Xhline{2\arrayrulewidth}
    \multicolumn{1}{p{1.5cm}|}{\centering $S_N + S_I$\\$\rightarrow T_I$} & \multicolumn{3}{p{3cm}}{\centering $N + M \rightarrow G$} & \multicolumn{3}{p{3cm}}{\centering $K + B \rightarrow W$} \\ \hline
    Metrics & P & R & F & P & R & F \\ \hline
    SB-TNT & $0.871$ & $0.932$ & $0.900$ & $0.543$ & $0.735$ & $0.624$ \\
    SB-DDAC & $0.871$ & $0.932$ & $0.901$ & $0.720$ & $0.784$ & $0.734$ \\
    SB-DDAS & $0.870$ & $0.932$ & $0.900$ & $0.639$ & $0.758$ & $0.671$ \\
    SB-STN & $0.931$ & $0.930$ & $0.931$ & $0.723$ & $0.764$ & $0.717$ \\
    STN & $0.951$ & $0.930$ & $0.938$ & $0.747$ & $0.793$ & $0.745$ \\
    CWAN & $0.871$ & $0.928$ & $0.898$ & $0.650$ & $0.772$ & $0.684$ \\ \hline
    \textbf{JSTN (Ours)} & $\textbf{0.954}$ & $\textbf{0.950}$ & $\textbf{0.952}$ & $\textbf{0.787}$ & $\textbf{0.808}$ & $\textbf{0.763}$ \\ \Xhline{2\arrayrulewidth}
    \end{tabular}}
  \label{tab:p_r_f1_performance_table}
  \vspace{-0.3cm}
\end{table}

\subsection{Performance Evaluation}\label{sec:section_performance_evaluation}

We firstly analyse the intrusion detection performance of the JSTN compared with other state-of-the-art counterparts on several randomly selected representative tasks. The evaluation results are presented in Table \ref{tab:giant_performance_table} - \ref{tab:p_r_f1_performance_table}. In Table \ref{tab:giant_performance_table}, the default ratio between the amount of labelled and unlabelled target II domain instances is set to $1:2$ when $T_{II} = B$, and is set to $1:5$ otherwise. As indicated in Table \ref{tab:giant_performance_table}, JSTN outperforms all other comparing methods over all tasks. Specifically, comparing with the double-source counterpart CWAN and STN, the best-performed single source method SB-STN, and the best traditional supervised ML method SVM, the JSTN yields a $9.42\%$, $8.02\%$, $7.75\%$ and $13.28\%$ performance boost, which is a significant improvement of detection accuracy. It is natural to observe this since although CWAN utilises both the source NI and II domain to facilitate intrusion detection for the target II domain, it does not specifically pay attention to semantic transfers such as the implicit or explicit semantics, which therefore verifies the usefulness of the robust semantic knowledge transfer utilised by the JSTN. Besides, the STN method does not consider the scenario semantics, which therefore results in hindered intrusion detection performance when huge heterogeneities present between NI and II domains. 

To verify the effectiveness of methods under varied $n_{TL}:n_{TU}$ ratios, especially under the extreme case where the amount of unlabelled target II domain data is significantly higher than the amount of labelled target II domain instances, $6$ tasks are randomly selected with the $n_{TL}:n_{TU}$ ratio varied between $1:2$ and $1:50$. Following \cite{10.1145/3394171.3413995,ning2021malware,yao2019heterogeneous}, the $1:50$ case is sufficient to represent an extremely label-scarce scenario. As shown in Table \ref{tab:ratio_performance_table1} and \ref{tab:ratio_performance_table2}, the JSTN outperforms all baseline methods by a large margin. Overall, the JSTN achieves a $7.36\%$, $6.6\%$, $5.77\%$ and $9.52\%$ amount of performance boost compared with the double-source method CWAN, STN, the best-performed single source baseline SB-STN, and the best-performed traditional supervised ML method SVM, respectively. Specifically, under the extreme $1:50$ case, the JSTN also shows robust performance, its performance only drops $1.13\%$ compared with the $1:10$ case while still outperforms the best DA method SB-STN and the best ML method NN by $5.57\%$ and $8.97\%$, which not only verifies the effectiveness of the JSTN when performing semantic knowledge transfer to facilitate intrusion detection in the IoT target domain, but also testifies the robustness of the JSTN when working on extremely scarcely-labelled target II domain. 

To further verify the efficacy of methods under evaluation metrics other than intrusion detection accuracy, we utilise precision, recall and F1-score as additional metrics and present the results on two randomly selected tasks as representatives in Table \ref{tab:p_r_f1_performance_table}. As we can see, the JSTN achieves the highest precision, recall and F1-score among these tasks over other DA-based baseline methods. By achieving the highest precision, it indicates that the JSTN model achieves the highest correctness among all network traffic that it flags as malicious attacks. Besides, achieving the best recall reveals the JSTN can detect most amount of malicious traffic out of all malicious behaviours, which indicates its effectiveness in terms of intrusion detection. Overall, the highest F1-score indicates the JSTN successfully balances between flagging as many intrusions from all malicious behaviours as possible, and meanwhile avoid triggering too many false alarms. Hence, together with the overall accuracy as indicated in Table \ref{tab:giant_performance_table} - \ref{tab:ratio_performance_table2}, the best performance on all evaluation metrics achieved by the JSTN model verifies its superiority. 

\begin{table}[!ht]
  \caption{Intrusion detection accuracy of the JSTN and its ablated counterparts. }
  \vspace{-0.3cm}
  \centering
  {\renewcommand{\arraystretch}{1.3}
  \begin{tabular}{c|ccc|c}
    \Xhline{2\arrayrulewidth}
    \multicolumn{1}{p{2.1cm}|}{\centering Ablated Method} & \multicolumn{1}{p{1cm}}{\centering $C + W$\\$\rightarrow B$} & \multicolumn{1}{p{1cm}}{\centering $K + M$\\$\rightarrow G$} & \multicolumn{1}{p{1cm}}{\centering $N + F$\\$\rightarrow B$} & \multicolumn{1}{p{0.6cm}}{\centering Avg} \\ \hline
    Full & $\textbf{69.69}$ & $\textbf{87.10}$ & $\textbf{86.94}$ & $\textbf{81.24}$ \\
    $\alpha = 0$ & $68.94$ & $86.42$ & $83.27$ & $79.54$ \\
    No WI & $67.96$ & $86.47$ & $85.14$ & $79.86$ \\
    $\beta = 0$ & $67.44$ & $86.29$ & $85.47$ & $79.73$ \\
    $\lambda = 0$ & $66.99$ & $86.58$ & $85.20$ & $79.59$ \\
    $\beta = \lambda = 0$ & $66.85$ & $86.19$ & $85.40$ & $79.48$ \\
    No PLR & $67.73$ & $86.60$ & $83.89$ & $79.41$ \\
    $\eta = 0$ & $65.40$ & $86.74$ & $84.18$ & $78.77$ \\
    $\gamma = 0$ & $66.57$ & $86.73$ & $85.17$ & $79.49$ \\
    $S_{NI}$ only & $67.18$ & $86.71$ & $84.93$ & $79.61$ \\
    $S_{II}$ only & $62.59$ & $85.34$ & $85.12$ & $77.68$ \\ \Xhline{2\arrayrulewidth}
    \end{tabular}}
  \label{tab:ablation_study_table}
  \vspace{-0.3cm}
\end{table}

\subsection{Ablation Study}\label{sec:section_ablation_study}

\begin{table*}[h]
  \caption{Intrusion detection performance when transferring semantic knowledge to facilitate target II domain intrusion detection via \RNum{1}: only a single source NI domain; \RNum{2}: two source NI domains; \RNum{3}: a source NI domain with a small-scale source II domain}
  \vspace{-0.3cm}
  \centering
  {\renewcommand{\arraystretch}{1.3}
  \begin{tabular}{c|cccccc}
    \Xhline{2\arrayrulewidth}
    Task & \multicolumn{1}{p{1.6cm}}{\centering $C + W \rightarrow B$} & \multicolumn{1}{p{1.6cm}}{\centering $N + F \rightarrow B$} & \multicolumn{1}{p{1.6cm}}{\centering $N + B \rightarrow W$} & \multicolumn{1}{p{1.6cm}}{\centering $K + W \rightarrow B$} & \multicolumn{1}{p{1.6cm}}{\centering $N + W \rightarrow G$} & \multicolumn{1}{p{0.6cm}}{\centering Avg} \\ \hline
    \centering \RNum{1}: $S_{NI} \rightarrow T_{II}$ & \multicolumn{1}{p{1.6cm}}{\centering $C \rightarrow B$\\$67.18$} & \multicolumn{1}{p{1.6cm}}{\centering $N \rightarrow B$\\$84.93$} & \multicolumn{1}{p{1.6cm}}{\centering $N \rightarrow W$\\$86.96$} & \multicolumn{1}{p{1.6cm}}{\centering $K \rightarrow B$\\$68.64$} & \multicolumn{1}{p{1.6cm}}{\centering $N \rightarrow G$\\$84.88$} & $78.52$ \\ \hline
    \multirow{2}{*}{\centering \RNum{2}: $S_{NI1} + S_{NI2} \rightarrow T_{II}$} & \multicolumn{1}{p{1.6cm}}{\centering $C + K \rightarrow B$\\$66.30$} & \multicolumn{1}{p{1.6cm}}{\centering $N + K \rightarrow B$\\$84.84$} & \multicolumn{1}{p{1.6cm}}{\centering $N + K \rightarrow W$\\$87.04$} & \multicolumn{1}{p{1.6cm}}{\centering $K + C \rightarrow B$\\$68.28$} & \multicolumn{1}{p{1.6cm}}{\centering $N + C \rightarrow G$\\$86.03$} & \multirow{2}{*}{\centering $78.70$} \\
     & \multicolumn{1}{p{1.6cm}}{\centering $C + N \rightarrow B$\\$68.91$} & \multicolumn{1}{p{1.6cm}}{\centering $N + C \rightarrow B$\\$84.87$} & \multicolumn{1}{p{1.6cm}}{\centering $N + C \rightarrow W$\\$87.08$} & \multicolumn{1}{p{1.6cm}}{\centering $K + N \rightarrow B$\\$67.86$} & \multicolumn{1}{p{1.6cm}}{\centering $N + K \rightarrow G$\\$85.78$} & \\ \hline
     \RNum{3}: $S_{NI} + S_{II} \rightarrow T_{II}$ & $\textbf{69.69}$ & $\textbf{86.94}$ & $\textbf{87.78}$ & $\textbf{69.73}$ & $\textbf{86.59}$ & $\textbf{80.15}$ \\ \Xhline{2\arrayrulewidth}
    \end{tabular}}
  \label{tab:ablation_study_scenario_semantic_table}
  \vspace{-0.3cm}
\end{table*}

\begin{figure*}[h]
  \begin{center}
    \includegraphics[width=0.8\textwidth,keepaspectratio]{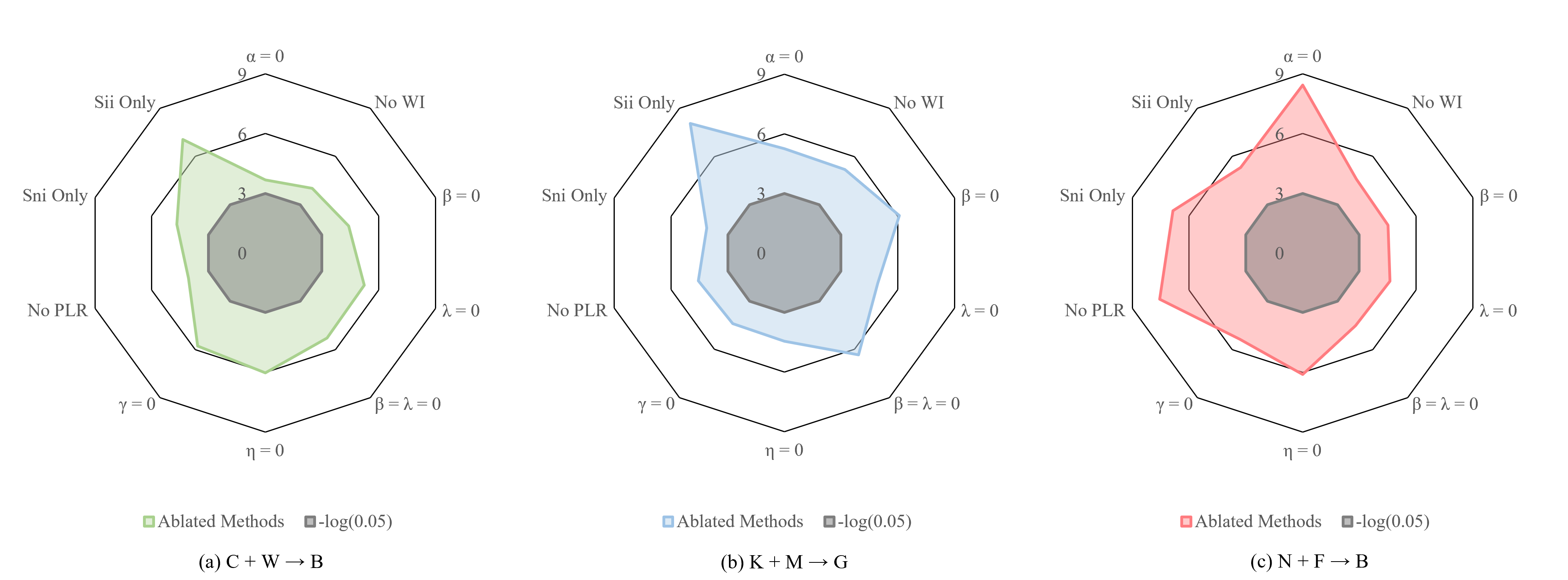}\\
    \caption{To verify the performance gains of the full JSTN over its ablated variants are statistically significant, significant T-tests with $0.05$ as the significant threshold are conducted on $3$ randomly picked tasks. The $-\log(0.05)$ significant threshold are indicated by the grey shaded area. In each dimension, the higher the value is, the more significant the performance gain is. }
    \vspace{-0.5cm}
    \label{fig:figure_ablation_study_significant_test}
  \end{center}
\end{figure*}

\textbf{Component Ablation Study} After performing the overall performance evaluation, we now verify the usefulness of each constituting semantic transfer component of the JSTN. The JSTN variants include the following: $(1)$ $\alpha = 0$, which ablates the weighted implicit semantic transfer; $(2)$ No WI, which turns off the weighting mechanism during implicit semantic transfer; $(3)$ $\beta = 0$, which removes the centroid-level explicit semantic alignment; $(4)$ $\lambda = 0$, which ablates the representative-level explicit semantic alignment; $(5)$ $\beta = \lambda = 0$, which completely turns off the hierarchical explicit alignment; $(6)$ No PLR, which assigns pseudo-label directly from the shared classifier $C$, without using the pseudo-label refiner $PLR$ that is geometric-aware; $(7)$ $\eta = 0$, which removes the categorical distribution preservation used during scenario semantic transfer; $(8)$ $\gamma = 0$, which turns off the domain discriminator $D$, part of the scenario semantic; $(9)$ $S_{NI}$ Only, which only uses a source NI domain, without considering the scenario semantic; $(10)$ $S_{II}$ Only, which only uses a source II domain, without utilising the knowledge rich source NI domain. 

The ablation performance on $3$ randomly selected representative tasks are indicated in Table \ref{tab:ablation_study_table}. The JSTN outperforms all its ablated variants, which verifies that all semantic transfer components are indispensable to facilitate a robust knowledge transfer. Without any constituting semantic, negative effects will be caused, which therefore leads to impaired intrusion detection performance. Among all these components, the weighted implicit semantic contributes around $1.7\%$ of performance improvement, while the weighting mechanism utilised during implicit semantic raises the performance by $1.38\%$. The hierarchical explicit semantic alignment contributes $1.76\%$ performance improvement on average. Specifically, only using the centroid-level or the representative-level explicit semantic alignment will cause the performance to drop by $1.51\%$ and $1.65\%$, respectively, which verify the importance of the hierarchical explicit semantic alignment. The pseudo-label refiner yields $1.83\%$ performance boost, which verifies the necessity to refine pseudo-labels. In terms of scenario semantic, using the domain discriminator will bring $1.75\%$ performance improvement. As part of the scenario semantic, the categorical distribution knowledge preservation yields $2.47\%$ performance increase. Additionally, the $S_{NI}$ only variant presents a performance reduction of around $1.63\%$ without the help of scenario semantic brought by the $S_{II}$ domain, and the $S_{II}$ only variant reduces the performance by $3.56\%$ due to lack of data and intrusion knowledge contained in it. The degraded performance of single domain variants further verifies the usefulness of the scenario semantic. 

\textbf{Necessity of Scenario Semantic} To have a closer look of the importance and necessity of the scenario semantic we proposed, detailed analyses are performed as indicated in Table \ref{tab:ablation_study_scenario_semantic_table}. When transferring knowledge to facilitate the intrusion detection of the target II domain, three variants are considered as follows: \RNum{1}, only transfer the knowledge via a single $S_{NI}$ domain; \RNum{2}, transfer the knowledge via two source NI domains $S_{NI1}$ and $S_{NI2}$, to make it a comparable \rw{replacement of} $S_{II}$, $S_{NI2}$ has the same scale as $S_{II}$; \RNum{3}, transfer the knowledge via the scenario semantic-enabled setting, i.e., a $S_{NI}$ domain facilitated with a small-scale $S_{II}$ domain. 

As we can observe from Table \ref{tab:ablation_study_scenario_semantic_table}, the scenario semantic-enabled variant achieves superior performance than other two variants by $1.63\%$ and $1.45\%$ on average, respectively. Hence, it indicates that a single source NI domain may present an overly large heterogeneous semantic gap, which will significantly hinder the intrusion detection performance without the help from the scenario semantic. Besides, by facilitating the source NI domain $S_{NI}$ with another NI domain, the semantic gap caused by domain heterogeneity is not effectively shortened, which is revealed by nearly the same performance between variant \RNum{1} and \RNum{2}. Furthermore, by leveraging a source II domain which is even $2$ - $6$ times smaller than the NI domain counterpart in scale, it can yield positive scenario semantic transfer to bridge the domain gap between heterogeneous NI and II domains, as verified by the superior performance. 

\textbf{Significant Test Verification} To verify the performance gains achieved by the full JSTN over its ablated variants are statistically significant, i.e., not observed randomly by chance, significant T-tests with $0.05$ as the significant threshold \rw{are performed} on $3$ randomly picked tasks, each is repeated $10$ times. The test results are illustrated in Figure \ref{fig:figure_ablation_study_significant_test}. The significant threshold $-\log(0.05)$ is indicated by the grey shaded area in the centre of each subfigure. Each dimension represents an ablated JSTN variant, the higher the value is, the more significant the performance gain is on this ablated component. As we can see, the full JSTN has a wider coverage on all dimensions under all tasks. A wider coverage than the grey shaded area indicates the test results among all ablated variants are significant. Hence, the usefulness and necessity of all constituting components of JSTN is verified with statistical significance. 

\subsection{Parameter Sensitivity Analysis}\label{sec:section_parameter_sensitivity_analysis}

\begin{figure*}[t]
  \begin{center}
    \includegraphics[width=\textwidth,keepaspectratio]{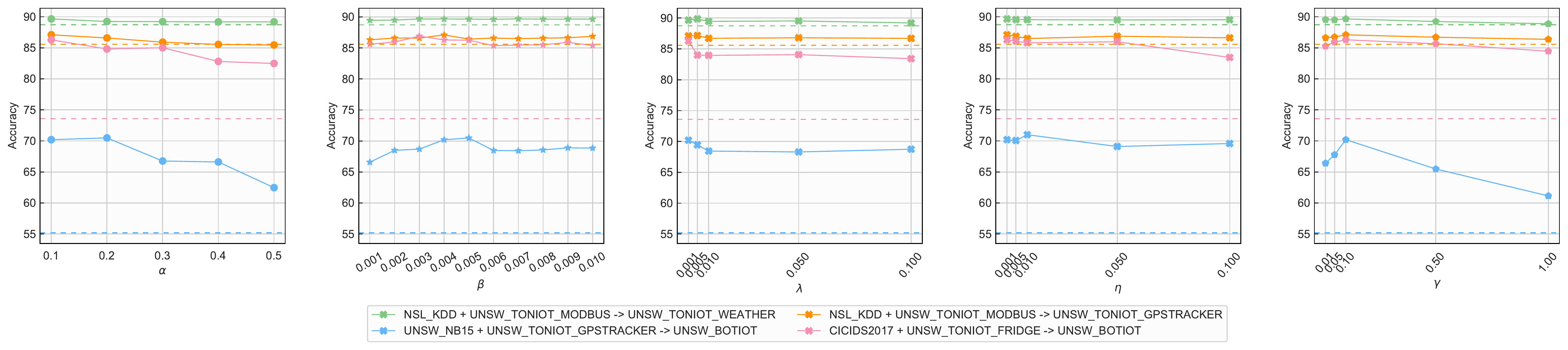}\\
    \caption{Parameter sensitivity of the JSTN for hyperparameter $\alpha$, $\beta$, $\lambda$, $\eta$ and $\gamma$ under their corresponding reasonable ranges on four randomly selected tasks. \rw{Each colour represents a task. Each colour has two lines, the solid line indicates the JSTN performance, and }the dashed \rw{horizontal line} indicates the accuracy of the best-performed baseline method of the corresponding task. }
    \vspace{-0.5cm}
    \label{fig:figure_parameter_sensitivity}
  \end{center}
\end{figure*}

\begin{table*}[!ht]
  \caption{\rw{Average training time per epoch (Measured in second, the lower the better). }}
  \vspace{-0.3cm}
  \centering
  \rw{{\renewcommand{\arraystretch}{1.3}
  \begin{tabular}{c|cccccc|c}
  \Xhline{2\arrayrulewidth}
  \multicolumn{1}{p{1.5cm}}{\centering $S_N + S_I$\\$\rightarrow T_I$} & \multicolumn{1}{p{1cm}}{\centering $C + W$\\$\rightarrow B$} & \multicolumn{1}{p{1cm}}{\centering $C + B$\\$\rightarrow G$} & \multicolumn{1}{p{1cm}}{\centering $N + G$\\$\rightarrow B$} & \multicolumn{1}{p{1cm}}{\centering $N + B$\\$\rightarrow W$} & \multicolumn{1}{p{1cm}}{\centering $K + W$\\$\rightarrow B$} & \multicolumn{1}{p{1cm}}{\centering $K + M$\\$\rightarrow G$} & \multicolumn{1}{p{0.6cm}}{\centering Avg} \\ \hline
  STN & $0.72$ & $0.75$ & $0.37$ & $0.38$ & $0.48$ & $0.34$ & $0.51$ \\
  CWAN & $1.10$ & $1.14$ & $0.59$ & $0.61$ & $0.76$ & $0.55$ & $0.79$ \\ \hline
  \textbf{JSTN (Ours)} & $\textbf{0.50}$ & $\textbf{0.51}$ & $\textbf{0.31}$ & $\textbf{0.36}$ & $\textbf{0.40}$ & $\textbf{0.30}$ & $\textbf{0.40}$ \\ \Xhline{2\arrayrulewidth}
  \end{tabular}}}
  \label{tab:computational_cost_per_epoch}
  \vspace{-0.3cm}
\end{table*}

\begin{table*}[!ht]
  \caption{\rw{Average inference time per unlabelled target instance (Measured in microsecond ($10^{-6}$ second), the lower the better). }}
  \vspace{-0.3cm}
  \centering
  \rw{{\renewcommand{\arraystretch}{1.3}
  \begin{tabular}{c|cccccc|c}
  \Xhline{2\arrayrulewidth}
  \multicolumn{1}{p{1.5cm}}{\centering $S_N + S_I$\\$\rightarrow T_I$} & \multicolumn{1}{p{1cm}}{\centering $C + W$\\$\rightarrow B$} & \multicolumn{1}{p{1cm}}{\centering $C + B$\\$\rightarrow G$} & \multicolumn{1}{p{1cm}}{\centering $N + G$\\$\rightarrow B$} & \multicolumn{1}{p{1cm}}{\centering $N + B$\\$\rightarrow W$} & \multicolumn{1}{p{1cm}}{\centering $K + W$\\$\rightarrow B$} & \multicolumn{1}{p{1cm}}{\centering $K + M$\\$\rightarrow G$} & \multicolumn{1}{p{0.6cm}}{\centering Avg} \\ \hline
  STN & $81.2$ & $52.6$ & $57.6$ & $55.1$ & $52.8$ & $58.8$ & $59.68$ \\
  CWAN & $56.9$ & $49.4$ & $78.1$ & $58.8$ & $65.0$ & $66.2$ & $62.40$ \\ \hline
  \textbf{JSTN (Ours)} & $\textbf{0.21}$ & $\textbf{0.22}$ & $\textbf{0.20}$ & $\textbf{0.21}$ & $\textbf{0.23}$ & $\textbf{0.18}$ & $\textbf{0.21}$ \\ \Xhline{2\arrayrulewidth}
  \end{tabular}}}
  \label{tab:computational_cost_per_instance_inferenced}
  \vspace{-0.3cm}
\end{table*}

To verify the parameter sensitivity of the JSTN model, we vary five major hyperparameters, i.e., $\alpha$, $\beta$, $\lambda$, $\eta$ and $\gamma$ within their corresponding reasonable value ranges. We randomly select four tasks as representatives and plot the results in Figure \ref{fig:figure_parameter_sensitivity}. The best-performed baseline method for each task is also plotted with the corresponding colour in dashed lines. As we can notice, \rw{when parameters vary, the performance of the JSTN model remains relatively stable without incurring severe fluctuation, as indicated by the relatively stable trend of each solid line. }Besides, \rw{the solid lines stay above their corresponding coloured dashed line in nearly all parameter ranges, which means the JSTN outperforms the corresponding best-performed counterpart under nearly all parameter ranges. Therefore, it verifies the robustness and effectiveness of the JSTN model. }

\subsection{\rw{Computational Efficiency}}\label{sec:section_computational_efficiency}

\rw{We further measure the computational efficiency of the JSTN model. The results on $6$ randomly selected representative tasks are presented in Table \ref{tab:computational_cost_per_epoch} for average training time per epoch, and in Table \ref{tab:computational_cost_per_instance_inferenced} for average inference time per unlabelled target instances. We only compare the JSTN with top-performed counterparts STN and CWAN. As indicated in Table \ref{tab:computational_cost_per_epoch}, the JSTN demonstrates the most efficient per-epoch training speed. The per-epoch performance boost achieved by JSTN is $21.6\%$ more efficient compared with the second-best performed counterpart STN. The most efficient per-epoch training speed reflects that the JSTN enjoys a relatively low computational complexity. The STN, CWAN and JSTN require $300$, $500$ and $1000$ epochs to train, therefore, the overall training time of these three methods stays comparable with each other. Given that the model training will be performed on relatively resource-rich devices, the computational cost of the JSTN model is satisfying. On the other hand, when performing the inference for unlabelled target intrusion data, the JSTN achieves the lowest per-instance inference time, as indicated in Table \ref{tab:computational_cost_per_instance_inferenced}. The JSTN even achieves $10^2$ times performance boost, thanks to the JSTN's excellent efficiency. Overall, the results verify the computational efficiency of the JSTN model. }

\section{Conclusion}\label{sec:section_conclusion}

In this paper, considering that the knowledge-rich network intrusion domain can facilitate more accurate intrusion detection for the data scarce IoT domain, we propose the JSTN network. Since there exists a significant semantic gap between NI and II domains due to heterogeneities, we utilise a small-scale auxiliary source II domain to endow the source NI domain with \rw{scenario semantics}. The categorical distribution knowledge is preserved between source domains, and the domain discriminator shortens the source domain gap, it also minimises the divergence between the whole source domain and the target domain, so that the source-target knowledge transfer effort will be eased. To preserve the categorical correlation enriched in the predicted distribution, we leverage the weighted implicit semantic transfer to achieve a more fine-grained knowledge learning and circumvent confounded categories for better discriminability. The implicit knowledge learning is guided by a weighting mechanism which depends on the divergence between each source domain and the target domain, so that well-learned source domain will be slightly suppressed while the more diverged source domain will be emphasised adaptively. Besides, we also tackle the problem from the distance perspective via the hierarchical explicit semantic alignment. Specifically, the centroid-level alignment achieves a more discriminative shared feature representation from a global perspective, while the representative-level alignment promotes better concentration during alignment and remains computation efficient. To better utilise unlabelled target II domain data while not suffering from the negative transfer brought by wrongly-assigned pseudo-labels, a geometric-aware pseudo-label refiner is used to boost the pseudo-label assignment confidence. By jointly utilising these three semantic transfer mechanisms, the JSTN model can learn a domain-invariant feature representation with fine-grained knowledge and high discriminability to facilitate more accurate IoT intrusion detection. Comprehensive experiments on several well-known ID datasets show the effectiveness of the JSTN compared with several state-of-the-art counterparts. The insight analyses also demonstrate the usefulness and necessity of each proposed semantics, which supports the joint semantic transfer of the JSTN. 

%%%%%%%%%%%%%%%%%%%%%%
%%%%%%%%%%%%%%%%%%%%%%
%% Acknowledgements %%
%%%%%%%%%%%%%%%%%%%%%%
%%%%%%%%%%%%%%%%%%%%%%

\ifCLASSOPTIONcompsoc
  \section*{Acknowledgments}
\else
  \section*{Acknowledgment}
\fi

This work is supported by Key-Area Research and Development Program of Guangdong Province (2020B010164002). 

\section*{Appendix}

\rw{The table below contains the notations used in this paper, and their corresponding interpretation. Note that the symbol $** \in \{SN, SI, TL, TU\}$, which stands for source network domain, source IoT domain, labelled target domain and unlabelled target domain, respectively. The $**$ carries the same meaning in the following notations. }

\begin{table}[!ht]
  \centering
  \caption{\rw{Notations and their corresponding interpretation. }}
  \label{tab:notation_table}
  \rw{\begin{tabular}{c|l}
    \hline
    Notation & \multicolumn{1}{c}{Interpretation} \\ \hline
    $\mathcal{D}_{**}$ & The intrusion domain \\
    $\mathcal{X}_{**}$ & The network traffic features of domain $**$ \\ 
    $\mathcal{Y}_{**}$ & The category labels of domain $**$ \\
    $x_{**_i}$ & The i\textsuperscript{th} instance of domain $**$ \\
    $y_{**_i}$ & The label of the i\textsuperscript{th} instance of domain $**$ \\
    $n_{**}$ & Number of instances in domain $**$ \\
    $d_{**}$ & The feature dimension of domain $**$ \\
    $K$ & Total number of categories \\
    $d_C$ & The dimension of the common feature subspace \\
    $f(x_i)$ & \multicolumn{1}{p{6.6cm}}{The feature representation of instance $x_i$ in the common feature subspace} \\
    $E_{**}(x_i)$ & The feature encoder for domain $**$ \\
    $\mathcal{L}_{SSD}$ & The scenario semantic loss yielded by the discriminator \\
    $C$ & The shared classifier \\
    $T_1$ & \multicolumn{1}{p{6.6cm}}{The temperature hyperparameter used in scenario semantic transfer} \\
    $\mathcal{X}_{**}^{(k)}$ & Class $k$ instances in domain $**$ \\
    $|\mathcal{X}_{**}^{(k)}|$ & Number of class $k$ instances in domain $**$ \\
    $q^{(k)}$ & \multicolumn{1}{p{6.6cm}}{The predicted probability distribution of class $k$ for the source NI domain} \\
    $p^{(k)}$ & \multicolumn{1}{p{6.6cm}}{The predicted probability distribution of class $k$ for the source II domain} \\
    $\mathcal{L}_{SSC}$ & \multicolumn{1}{p{6.6cm}}{The Scenario semantic loss yielded by distribution matching} \\
    $T_2$ & \multicolumn{1}{p{6.6cm}}{The temperature hyperparameter used in weighted implicit semantic transfer} \\
    $\mathcal{L}_{sf}^{S*}$ & \multicolumn{1}{p{6.6cm}}{The soft loss based on source domain $S*$ in the weighted implicit semantic transfer, $S* \in \{SN, SI\}$} \\
    $\mathcal{L}_{hd}$ & \multicolumn{1}{p{6.6cm}}{The hard loss of target TL domain in the weighted implicit semantic transfer} \\
    $\mathcal{L}_{ce}$ & The cross entropy loss \\
    $\mu_{S*}^{(k)}$ & The class $k$ centroid of source domain $S* \in \{SN, SI\}$ \\
    $d_{<SN, TL>}$ & \multicolumn{1}{p{6.6cm}}{The divergence between source NI domain and labelled target II domain} \\
    $\omega_{<SN, TL>}$ & \multicolumn{1}{p{6.6cm}}{The weight for source NI domain in the weighted implicit semantic transfer} \\ 
    $\mathcal{L}_{WIS}$ & The weighted implicit semantic transfer loss \\ 
    $y_i^{<NN>}$ & The neural network label for unlabelled target instance $x_i$ \\ 
    $\mu^{(k)}$ & \multicolumn{1}{p{6.6cm}}{The class $k$ centroid of all labelled source and target instances} \\ 
    $y_i^{<GS>}$ & \multicolumn{1}{p{6.6cm}}{The geometric similarity-based label for unlabelled target instance $x_i$} \\ 
    $\mathcal{X}_{T}$ & \multicolumn{1}{p{6.6cm}}{The features of target domain combined with labelled and pseudo-labelled target instances} \\ 
    $\mathcal{Y}_{T}$ & \multicolumn{1}{p{6.6cm}}{The labels of target domain combined with labelled and pseudo-labelled target instances} \\ 
    $\hat{x_i}$ & Pseudo-labelled unlabelled target instance $x_i$ \\ 
    $\hat{y_i}$ & \multicolumn{1}{p{6.6cm}}{The assigned pseudo-label for unlabelled target instance $x_i$} \\ 
    $\mathcal{L}_{ESC}$ & The global centroid-level explicit semantic transfer loss \\ 
    $\mathcal{L}_{ESR}$ & \multicolumn{1}{p{6.6cm}}{The local representative-level explicit semantic transfer loss} \\ 
    $R$ & Number of representatives selected for each category \\ 
    $r_{S}^{(k)}(i)$ & \multicolumn{1}{p{6.6cm}}{The i\textsuperscript{th} class $k$ representative of the source domain S ($S = SN \cup SI$)} \\ 
    $\mathcal{L}_{sup}$ & \multicolumn{1}{p{6.6cm}}{The supervision loss of the source domain S ($S = SN \cup SI$)} \\ 
    $\beta, \lambda, \gamma, \eta$ & \multicolumn{1}{p{6.6cm}}{The hyperparameters that balance $\mathcal{L}_{ESC}$, $\mathcal{L}_{ESR}$, $\mathcal{L}_{SSD}$ and $\mathcal{L}_{SSC}$, respectively. }\\
    $TP^{(k)}$ & True positive value for category $k$ \\ \hline
  \end{tabular}}
\end{table}

\ifCLASSOPTIONcaptionsoff
  \newpage
\fi

%%%%%%%%%%%%%%%%%%%%%%%%%%%%%
%%%%%%%%%%%%%%%%%%%%%%%%%%%%%
%% Bibliography References %%
%%%%%%%%%%%%%%%%%%%%%%%%%%%%%
%%%%%%%%%%%%%%%%%%%%%%%%%%%%%

\bibliographystyle{IEEEtran}
\bibliography{JSTN}

%\newpage

%%%%%%%%%%%%%%%
%%%%%%%%%%%%%%%
%% Biography %%
%%%%%%%%%%%%%%%
%%%%%%%%%%%%%%%

\vspace{-0.8cm}
\begin{IEEEbiography}[{\includegraphics[width=1in,height=1.25in,clip,keepaspectratio]{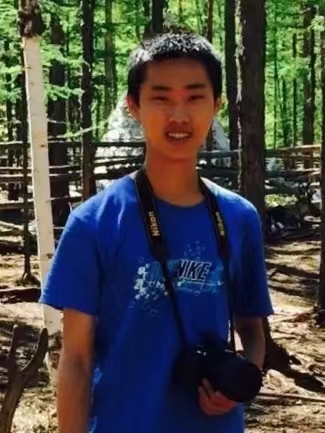}}]{Jiashu Wu} received BSc. degree in Computer Science and Financial Mathematics \& Statistics from the University of Sydney, Australia (2018), and M.IT. degree in Artificial Intelligence from the University of Melbourne, Australia (2020). He is currently pursuing his Ph.D. at the University of Chinese Academy of Sciences (Shenzhen Institute of Advanced Technology, Chinese Academy of Sciences). His research interests including machine learning and cloud computing. 
\end{IEEEbiography}
\vspace{-0.8cm}
\begin{IEEEbiography}[{\includegraphics[width=1in,height=1.25in,clip,keepaspectratio]{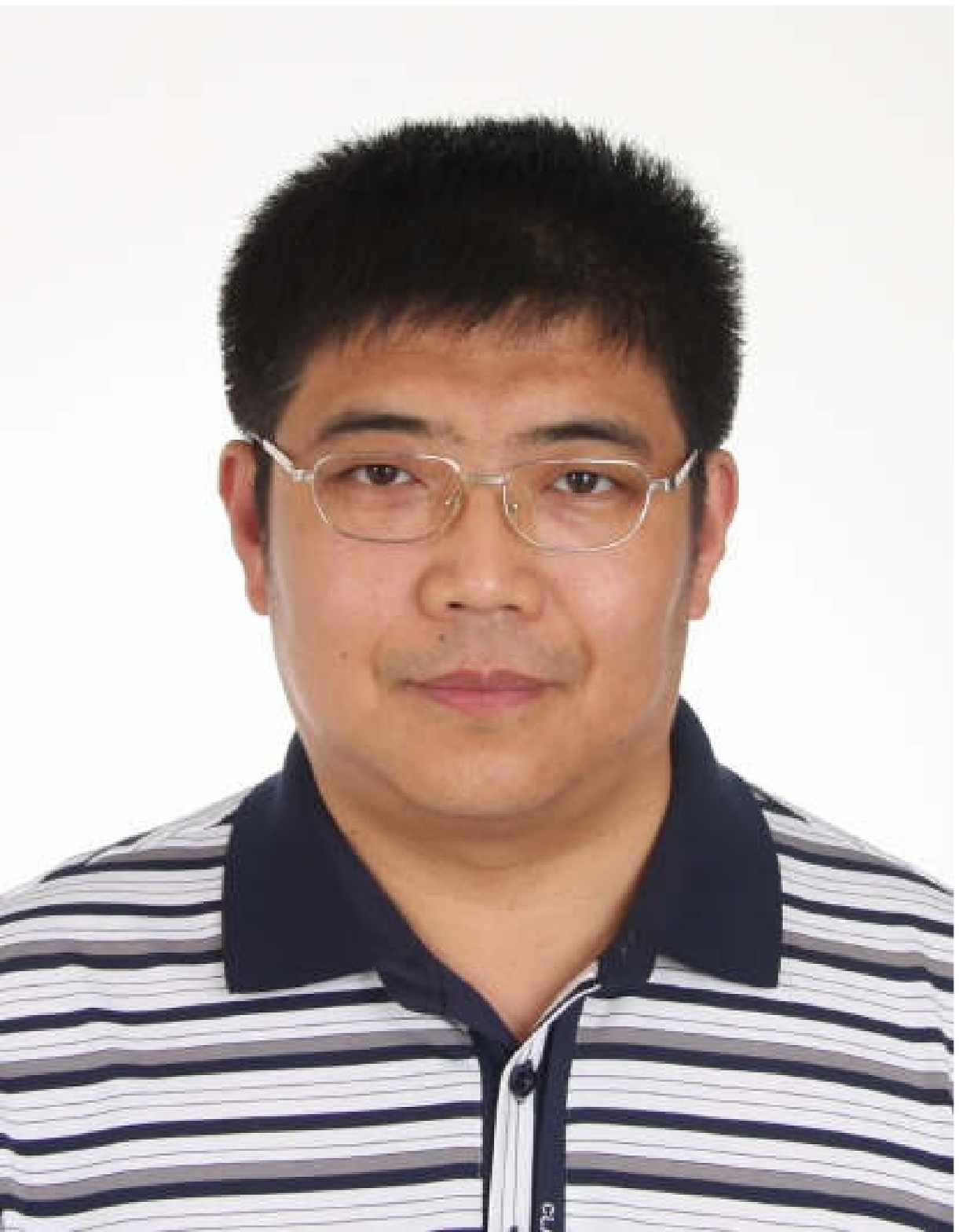}}]{Yang Wang} received the BSc. degree in applied mathematics from Ocean University of China (1989), and the MSc. and Ph.D. degrees in computer science from Carlton University (2001) and University of Alberta, Canada (2008), respectively. He is currently with Shenzhen Institutes of Advanced Technology, Chinese Academy of Sciences, as a professor and with Xiamen University as an adjunct professor. His research interests include service and cloud computing, programming language implementation, and software engineering. He is an Alberta Industry R\&D Associate (2009-2011), and a Canadian Fulbright Scholar (2014-2015). 
\end{IEEEbiography}
\vspace{-0.8cm}
\begin{IEEEbiography}[{\includegraphics[width=1in,height=1.25in,clip,keepaspectratio]{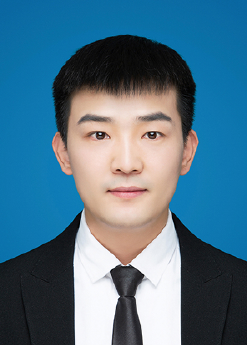}}]{Binhui Xie} is a Ph.D. student at the School of Computer Science and Technology, Beijing Institution of Technology. His research interests focus on computer vision and transfer learning.
\end{IEEEbiography}
\vspace{-0.8cm}
\begin{IEEEbiography}[{\includegraphics[width=1in,height=1.25in,clip,keepaspectratio]{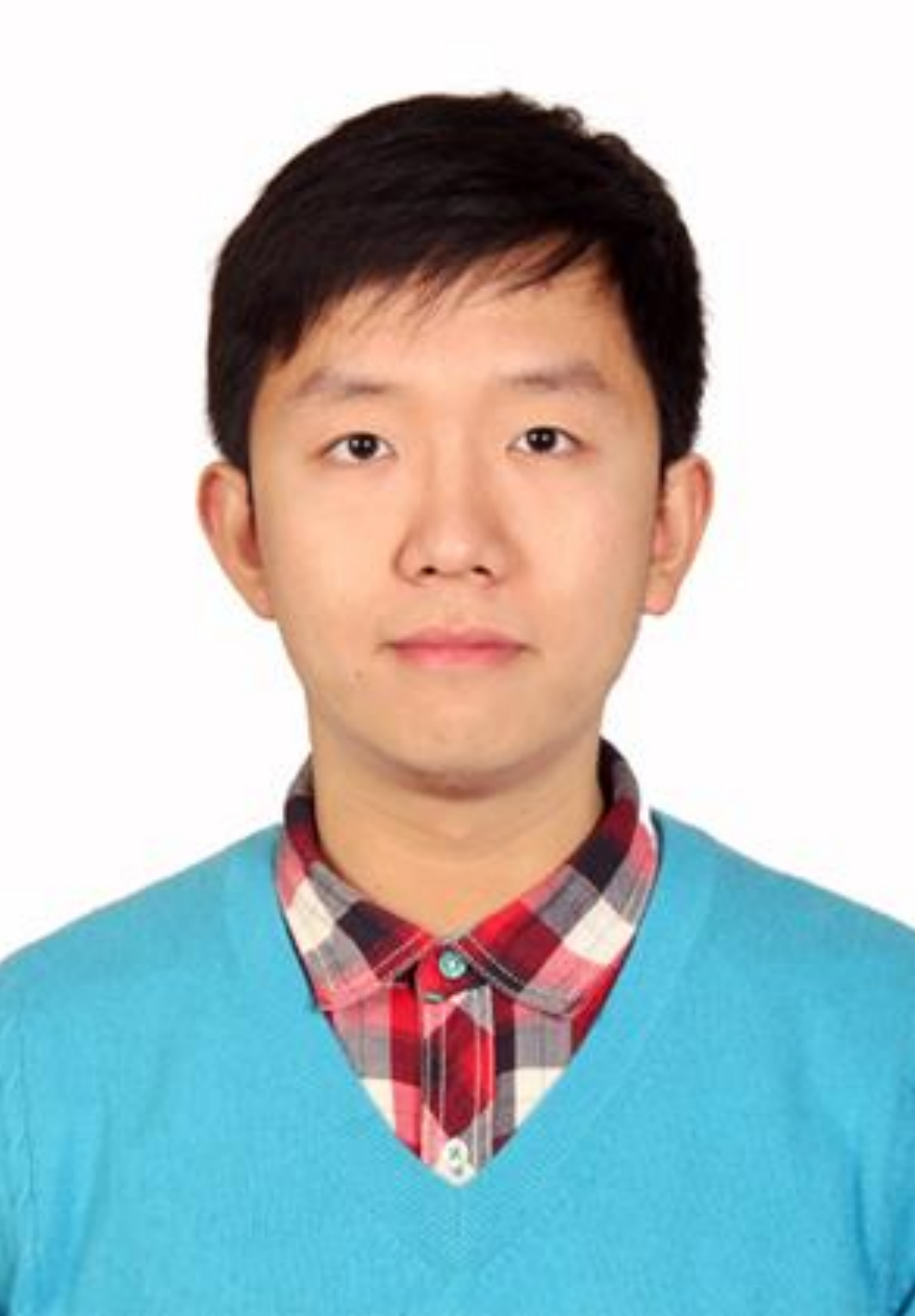}}]{Shuang Li} received the Ph.D. degree in control science and engineering from the Department of Automation, Tsinghua University, Beijing, China, in 2018. He was a Visiting Research Scholar with the Department of Computer Science, Cornell University, Ithaca, NY, USA, from November 2015 to June 2016. He is currently an Assistant Professor with the school of Computer Science and Technology, Beijing Institute of Technology, Beijing. His main research interests include machine learning and deep learning, especially in transfer learning and domain adaptation.
\end{IEEEbiography}
\vspace{-0.8cm}
\begin{IEEEbiography}[{\includegraphics[width=1in,height=1.25in,clip,keepaspectratio]{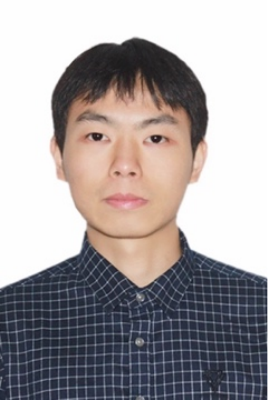}}]{Hao Dai} received the BSc. and M.Sc degrees in Communication and Electronic Technology from the Wuhan University of Technology in 2015 and 2017, respectively. He is currently working toward the Ph.D. degree in the Shenzhen Institute of Advanced Technology, Chinese Academy of Sciences. His research interests include mobile edge computing, federated learning and deep reinforcement learning. 
\end{IEEEbiography}
\vspace{-0.8cm}
\begin{IEEEbiography}[{\includegraphics[width=1in,height=1.25in,clip,keepaspectratio]{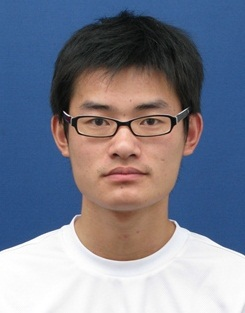}}]{Kejiang Ye}
received his BSc. and Ph.D. degree in Computer Science from Zhejiang University in 2008 and 2013 respectively. He was also a joint PhD student at University of Sydney from 2012 to 2013. After graduation, he worked as Post-Doc Researcher at Carnegie Mellon University from 2014 to 2015 and Wayne State University from 2015 to 2016. He is currently an Associate Professor at Shenzhen Institutes of Advanced Technology, Chinese Academy of Science. His research interests focus on the performance, energy, and reliability of cloud computing and network systems.
\end{IEEEbiography}
\vspace{-0.8cm}
\begin{IEEEbiography}[{\includegraphics[width=1in,height=1.25in,clip,keepaspectratio]{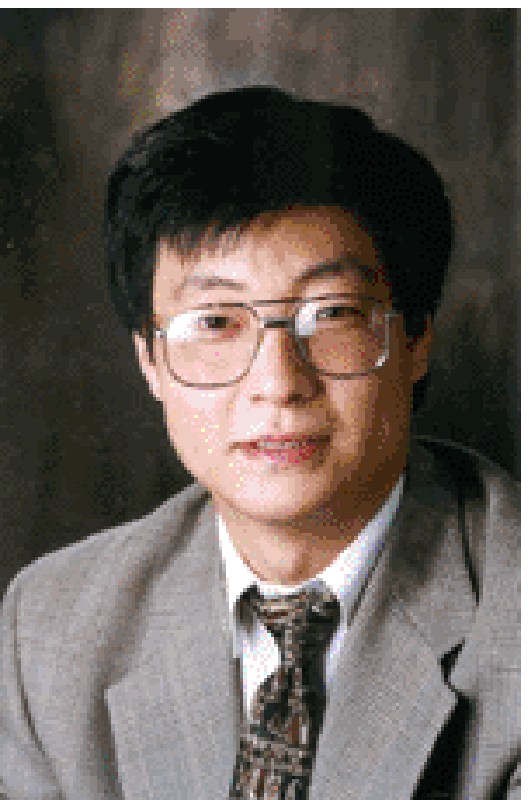}}]{Chengzhong Xu} received the Ph.D. degree from the University of Hong Kong in 1993. He is currently the Dean of Faculty of Science and Technology, University of Macau, China, and the Director of the Institute of Advanced Computing and Data Engineering, Shenzhen Institutes of Advanced Technology of Chinese Academy of Sciences.His research interest includes parallel and distributed systems, service and cloud computing, and software engineering. He has published more than 200 papers in journals and conferences. He serves on a number of journal editorial boards, including IEEE TC, IEEE TPDS, IEEE TCC, JPDC and China Science Information Sciences. He is a fellow of the IEEE.
\end{IEEEbiography}

\end{document}